\documentclass[twocolumn,prl,superscriptaddress,twocolumn]{revtex4-2}
\usepackage[utf8]{inputenc}
\usepackage{color}
\usepackage{amsmath}
\usepackage{amssymb}
\usepackage{graphicx}
\usepackage{booktabs}
\usepackage[unicode=true,
 bookmarks=true,bookmarksnumbered=false,bookmarksopen=false,
 breaklinks=false,pdfborder={0 0 1},backref=false,colorlinks=true]
 {hyperref}
\hypersetup{
 linkcolor=magenta, urlcolor=blue, citecolor=blue, pdfstartview={FitH}, unicode=true}
\usepackage{booktabs}

\makeatletter

\newcommand{\diff}[1]{\textcolor{black}{#1}}

\usepackage{amsfonts}\usepackage{tabularx}\usepackage{dcolumn}\usepackage{bm}\usepackage{epstopdf}
\usepackage{times}
\hypersetup{urlcolor=blue}

\usepackage{listings}
\usepackage{braket}
\usepackage{algorithm}  
\usepackage{algorithmicx}  
\usepackage{algpseudocode}  
\makeatother

\begin{document}

\title{\diff{Adversarial} Machine Learning Phases of Matter}

\author{Si Jiang}\thanks{These authors contributed equally to this work.}
\affiliation{Center for Quantum Information, IIIS, Tsinghua University, Beijing
100084, People\textquoteright s Republic of China}
\author{Sirui Lu}\thanks{These authors contributed equally to this work.}
\affiliation{Center for Quantum Information, IIIS, Tsinghua University, Beijing
100084, People\textquoteright s Republic of China}
\affiliation{Max-Planck-Institut f\"ur Quantenoptik, Hans-Kopfermann-Str.\ 1, D-85748 Garching, Germany}
\author{Dong-Ling Deng}\email{dldeng@tsinghua.edu.cn}
\affiliation{Center for Quantum Information, IIIS, Tsinghua University, Beijing
100084, People\textquoteright s Republic of China}
\affiliation{Shanghai Qi Zhi Institute, 41th Floor, AI Tower, No. 701 Yunjin Road, Xuhui District, Shanghai 200232, China}

\begin{abstract}
\diff{We study the robustness of machine learning approaches to adversarial perturbations, with a focus on supervised learning scenarios. } We find that typical phase classifiers based on deep neural networks are extremely vulnerable to adversarial perturbations: adding a tiny amount of carefully crafted noises into the original legitimate examples will cause the classifiers to make incorrect predictions at a notably high confidence level. \diff{ Through the lens of activation maps, we find that some important underlying physical principles and symmetries remain to be adequately captured for classifiers with even near-perfect performance. This explains why adversarial perturbations exist for fooling these classifiers. In addition, we find that, after adversarial training the classifiers will become more consistent with physical laws and consequently more robust to certain kinds of adversarial perturbations.
 Our results provide valuable guidance for both theoretical and experimental future studies on applying machine learning techniques to condensed matter physics.}
\end{abstract}

\maketitle
Machine learning is currently revolutionizing many technological areas
of modern society, ranging from image/speech recognition to content
filtering on social networks and self-driving cars \cite{Lecun2015Deep,Jordan2015Machine}.
Recently, its tools and techniques have been adopted to tackle intricate
quantum many-body problems \cite{Sarma2019Machine,Carleo2016Solving,Torlai2018Neural,
Nomura2017Restricted,You2017Machine,Deng2017Machine,Deng2017MachineBN,
Deng2017Quantum,Gao2017Efficient,melko2019restricted,Chng2017Machine,Wang2016Discovering},
where the exponential scaling of the Hilbert space dimension poses
a notorious challenge. In particular, a number of supervised and unsupervised
learning methods have been exploited to classify phases of matter
and identify phase transitions \cite{Wang2016Discovering,Zhang2016Triangular,Carrasquilla2017Machine,van2017Learning,
Broecker2017Machine,Chng2017Machine,Wetzel2017Unsupervised,Hu2017Discovering,
Hsu2018Machine,rodriguez2019identifying,Zhang2018Machine,
Huembeli2018Identifying,Suchsland2018Parameter,ohtsuki2016deep,ohtsuki2017deep,ohtsuki2019drawing,
greplova2019unsupervised,Hernandez2018Extrapolating}.
Following these approaches, notable proof-of-principle experiments
with different platforms \cite{Lian2019Machine,rem2019identifying,bohrdt2019classifying,zhang2019machine},
including electron spins in diamond nitrogen-vacancy (NV) centers \cite{Lian2019Machine},
doped $\text{CuO}_{2}$ \cite{zhang2019machine}, and cold atoms in
optical lattices \cite{rem2019identifying,bohrdt2019classifying},
have also been carried out subsequently, showing great potentials
for unparalleled advantages of machine learning approaches compared
to traditional means. 

\begin{figure}
\includegraphics[width=0.47\textwidth]{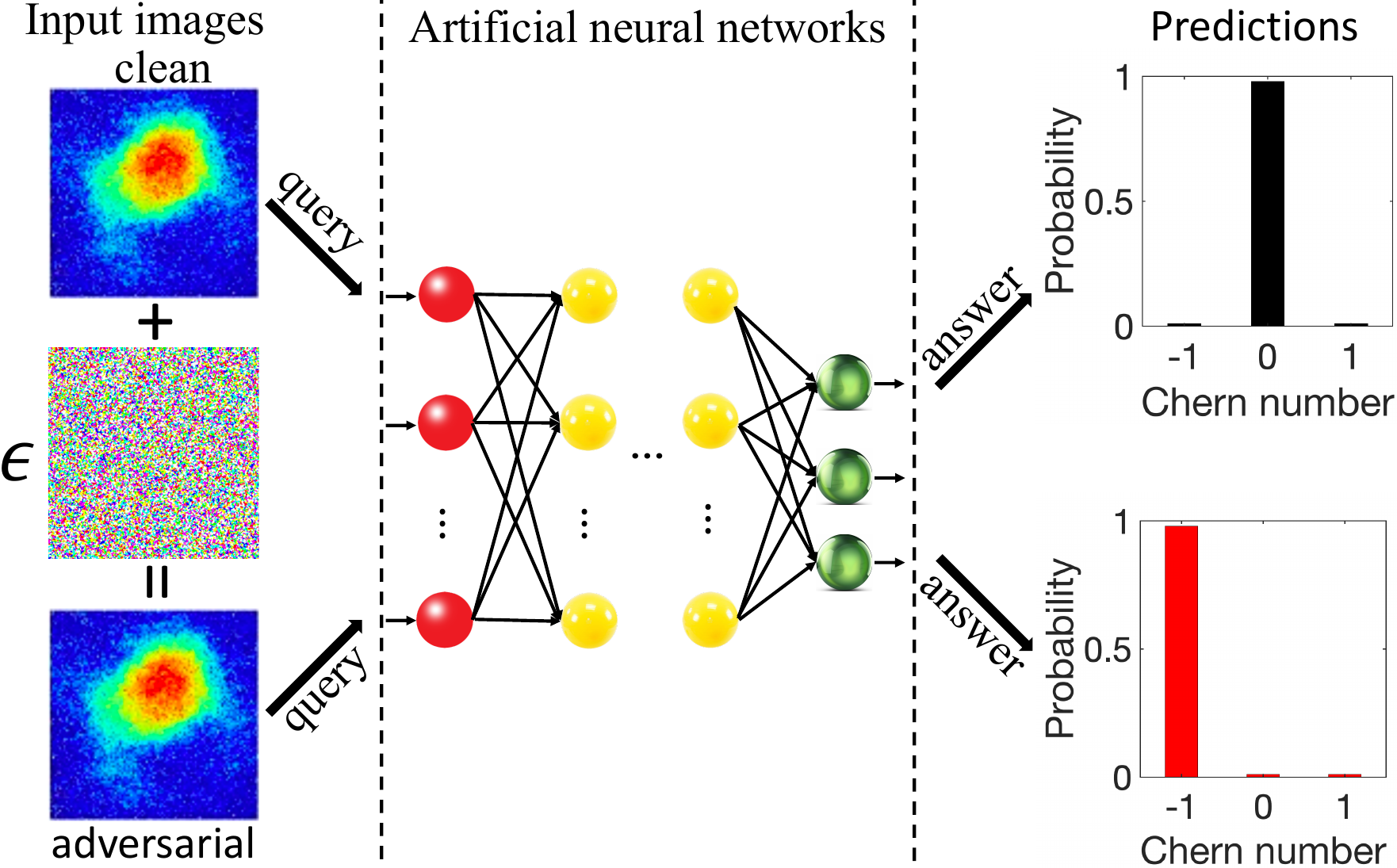}
\caption{A schematic illustration for the vulnerability of machine learning
phases of matter. For a clean image, such as the time-of-flight image
obtained in a recent cold-atom experiment \cite{rem2019identifying},
a trained neural network (i.e., the classifier) can successfully predict
its corresponding Chern number with nearly unit accuracy. However,
if we add a tiny adversarial perturbation (which is imperceptible
to human eyes) to the original image, the same classifier will misclassify
the resulted image into an incorrect category with nearly unit confidence. \label{fig:A-schematic-illustration}}
\end{figure}

An important question of both theoretical and experimental relevance
concerns the reliability of such machine-learning approaches to condensed
matter physics: are these approaches robust to adversarial perturbations,
which are deliberately crafted in a way intended to fool the classifiers?
In the realm of adversarial machine learning \cite{biggio2018wild,Huang2011Adversarial,vorobeychik2018adversarial,miller2019adversarial,goodfellow2014explaining,liu2019vulnerability,Schmidt2018Adv},
it has been shown that machine learning models can be surprisingly
vulnerable to adversarial perturbations if the dimension of the data
is high enough\textemdash one can
often synthesize small, imperceptible perturbations of the input data
to cause the model make highly-confident but erroneous predictions.
A prominent adversarial example that clearly manifests such vulnerability
of classifiers based on deep neural networks was first observed by
Szegedy \textit{et al.} \cite{Szegedy2013Intriguing}, where adding
a small adversarial perturbation, although unnoticeable to human eyes,
will cause the classifier to miscategorize a panda as a gibbon with
confidence larger than $99\%$. In this paper, we investigate the
vulnerability of machine learning approaches in the context of classifying
different phases of matter, with a focus on supervised
learning based on deep neural networks (see Fig. \ref{fig:A-schematic-illustration}
for an illustration).

We find that typical phase classifiers based on deep neural networks
are likewise extremely vulnerable to adversarial perturbations. This
is demonstrated through two concrete examples, which cover different
phases of matter (including both symmetry-breaking and symmetry-protected
topological phases) and different strategies to
obtain the adversarial perturbations. \diff{To better understand why these adversarial examples can fool the classifier \diff{in the physics context}, we open up the neural network and use an idea borrowed from the machine learning community, called activation map \cite{zhou2015object, zhou2015learning}, to study how the classifier \diff{infers} 
different phases of matter.} \diff{Further, we show that an adversarial training-based defense strategy improves classifiers' ability to resist specific perturbations and how well the underlying physical principles are captured.
Our results shed light on the fledgling field of machine-learning applications in condensed matter physics, which may provide an important paradigm 
for future theoretical and experimental studies as the field matures.}

\diff{To begin with, we introduce the main ideas of adversarial machine learning, which involves the generation and defense of adversarial examples \cite{biggio2018wild,Huang2011Adversarial,vorobeychik2018adversarial,miller2019adversarial}. \diff{Adversarial examples are instances with small intentionally crafted perturbations to cause the classifier make incorrect predictions.}
\diff{Under the }supervised learning scenario, we have a training data set with labels $\mathcal{D}_{n}=\{(\boldsymbol{x}^{(1)},y^{(1)}),\cdots,(\boldsymbol{x}^{(n)},y^{(n)})\}$, a classifier $h(\cdot;\theta)$ and a loss function $L$ to evaluate the classifier's performance. Adversarial examples generation task can be reduced to an optimization problem: searching for a bounded perturbation that maximizes the loss function \cite{VMLPMSup}}:
\begin{eqnarray}
\max_{\delta\in\Delta} & \;L(h(\boldsymbol{x}^{(i)}+\delta;\theta),y^{(i)}).\label{eq:AdvLmaxLoss}
\end{eqnarray}
\diff{In the machine learning literature, a number of methods have been proposed to solve the above optimization problem, along with corresponding defense strategies  \cite{storn1997differential,das2010differential,goodfellow2014explaining,Madry2017Towards,dong2018boosting}. We employ some of these methods and one general defense strategy, adversarial training, on two concrete examples: one concerns the conventional paramagnetic/ferromagnetic phases with a two-dimensional classical Ising model \cite{Carrasquilla2017Machine, Wang2016Discovering, van2017Learning}; the other involves topological phases with experimental raw
data generated by a solid-state quantum simulator~\cite{Lian2019Machine}. }

\textit{The ferromagnetic Ising model.}\textemdash The first example
we consider involves the ferromagnetic Ising model defined
on a 2D square lattice: $H_{\text{Ising}}  =  -J\sum_{\langle ij\rangle}\sigma_{i}^{z}\sigma_{j}^{z}$,
where the Ising variables $\sigma_{i}^{z}=\pm1$ and the coupling
strength $J\equiv1$ is set to be the energy unit. This model features
a well-understood phase transition at the critical temperature $T_{c}=2/\ln(1+\sqrt{2})\approx2.366$
\cite{Onsager1944Crystal}, between a high-temperature paramagnetic
phase and a low-temperature ferromagnetic phase. 
In the context of machine learning phases of matter, different pioneering
approaches, including these based on supervised learning \cite{Carrasquilla2017Machine},
unsupervised learning~\cite{Wang2016Discovering}, or a confusion
scheme combining both~\cite{van2017Learning},
have been introduced  to classify the ferromagnetic/paramagnetic
phases hosted by the above 2D Ising model. In particular, Carrasquilla
and Melko first explored a supervised learning scheme based on a fully
connected feed-forward neural network \cite{Carrasquilla2017Machine}.
They used equilibrium spin configurations sampled from Monte Carlo
simulations to train the network and demonstrated that after training
it can correctly classify new samples with notably high accuracy.
Moreover, through scanning the temperature the network can also locate
the the transition temperature $T_{c}$ and extrapolate the critical
exponents that are crucial in the study of phase transitions.

\begin{figure}[ht]
\includegraphics[width=0.49\textwidth]{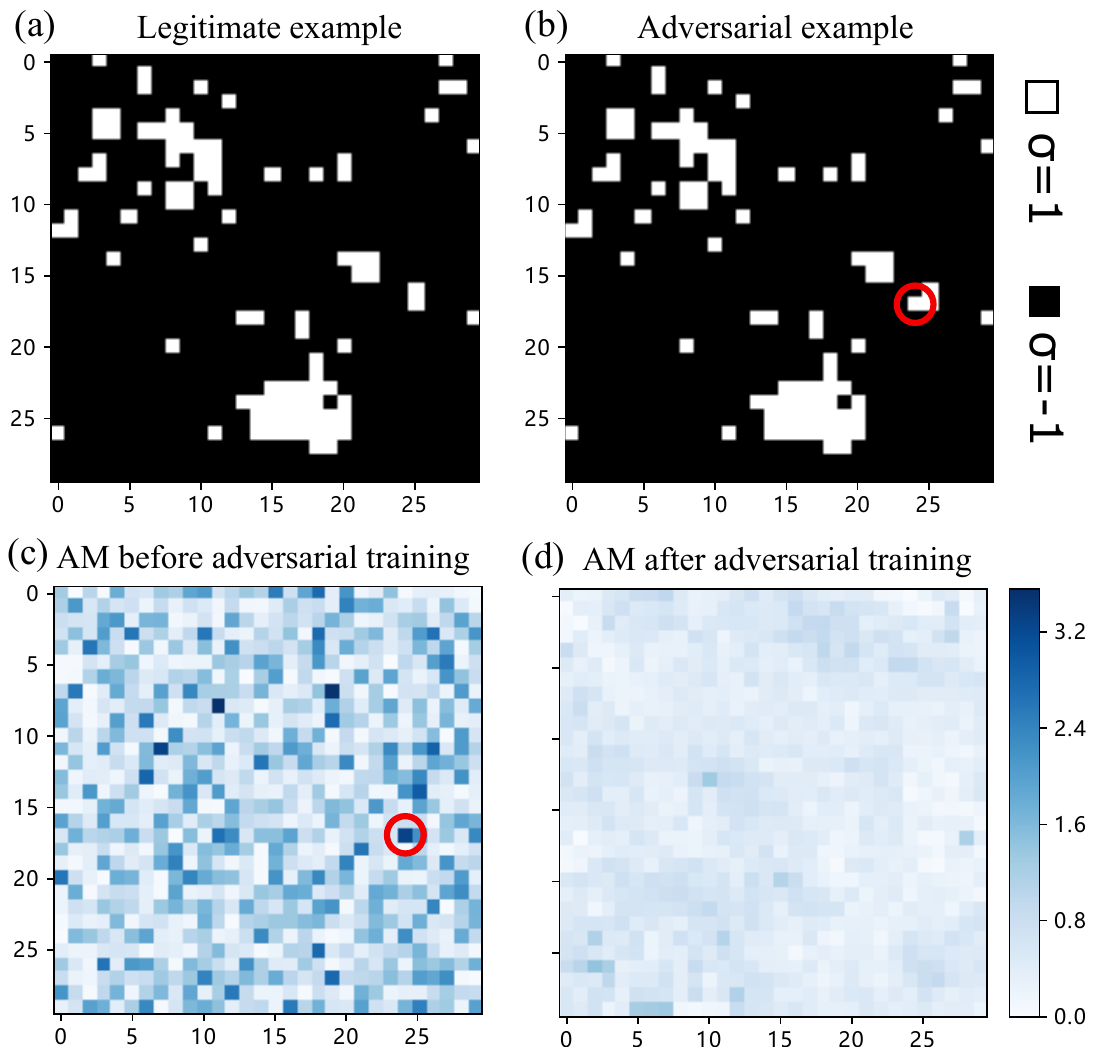}

\caption{\diff{(a) A legitimate sample of the spin configuration in the ferromagnetic phase with $M=|\sum_i^N\sigma_i|/N=0.791$. (b) An adversarial example obtained by the differential evolution algorithm (DEA), which only differs from the original legitimate one by flipping one spin (in red circle). (c) The activation map (AM) of the original classifier. The spins at positions with darker colors contribute more to the confidence of being ferromagnetic phase. (d) The activation map of the classifier after adversarial training. The map becomes much flatter and the variance of each position's activation value becomes much smaller.}
\label{fig:IsingM}}
\end{figure}

\diff{
To study the robustness of these introduced machine learning approaches to adversarial perturbations, we first train a powerful classifier which has comparable performance with the ones shown in \cite{Carrasquilla2017Machine}. 
After training, the network can successfully classify data from the test set with a high accuracy larger than $97\%$ \cite{VMLPMSup}. Then we try to obtain adversarial perturbations to attack this seemingly ideal classifier. It is natural to consider discrete attack in this scenario since the spin configuration in Ising model can be only discretely changed as spin flips. We apply the differential evolution algorithm (DEA) \cite{su2019one} 
to the Monte Carlo sampled spin configurations and obtain the corresponding adversarial perturbations.} \diff{A concrete example found by DEA is illustrated in Fig. \ref{fig:IsingM}(a-b). Initially, the legitimate example shown in (a) is in the ferromagnetic phase, which has magnetization $M=|\sum_i^N\sigma_i|/N=0.791$ and the classifier classifies it into the correct phase with confidence $72\%$. DEA obtains an adversarial example shown in (b) by flipping only a single spin, which is located in the red circle. This new spin configuration has almost the same magnetization $M=0.789$ and should still belong to the ferromagnetic phase, but the classifier misclassifies it into the paramagnetic phase with confidence $60\%$.} If we regard $H_{\text{Ising}}$ as a quantum Hamiltonian and allow the input data to be continuously modified, one can also consider a continuous attack scenario and obtain various adversarial examples, as shown in the Supplementary Material \cite{VMLPMSup}.

\diff{To understand why this adversarial example \diff{crafted with tiny changes leads to misclassification,}
we dissect the classifier by estimating each position's importance to the final prediction, which we call the activation map of the classifier \cite{VMLPMSup}. In Fig. \ref{fig:IsingM}(c) we depict the activation map for ferromagnetic phase. \diff{It is evident that} the classifier makes prediction mainly based on positions with large activation values (dark colors). The position where \diff{the adversarial} spin flip happens in Fig.~\ref{fig:IsingM}(b) has an activation value $3.28$, which is the forth largest among all 900 positions. \diff{Then we enumerate} 
all positions with activation values larger than $2.6$ and find that single spin flips, which changes the contribution to ferromagnetic phase from positive to negative, can all lead to misclassification \cite{VMLPMSup}. 
The values at different positions for the activation map is \diff{found to be }nonuniform, which contradicts to the 
\diff{physical knowledge} that each spin contributes equally to the order parameter $M$. 
This explains why the classifier is vulnerable to these particular spin flips. 
 We remark that in the traditional machine learning realm of classifying daily-life images (such as images of cats and dogs), such an explanation is unattainable due to the absence of a sharply defined ``order parameter".  }

\textit{Topological phases of matter.}\textemdash 
Unlike conventional phases
(such as the paramagnetic/ferromagnetic phases discussed above), topological
phases do not fit into the paradigm of symmetry breaking \cite{lifshitz2013statistical}
and are described by nonlocal topological invariants \cite{qi2011topological,hasan2010colloquium},
rather than local order parameters. This makes topological phases
harder to learn in general. Notably, a number of different approaches,
based on either supervised \cite{Zhang2016Triangular,Zhang2017Machine,Zhang2018Machine}
or unsupervised \cite{rodriguez2019identifying, Yu2021Unsupervised, Scheurer2020Unsupervised, Long2020Unsupervised, Lidiak2020Unsupervised, Che2020Topological, fukushima2019featuring, Schafer2019Vector, Balabanov2020Unsupervised, Alexandrou2020The, Greplov2019nsupervised, Arnold2021Interpretable} learning paradigms,
have been proposed recently and some of them been demonstrated in
proof-of-principle experiments \cite{Lian2019Machine,rem2019identifying}. 

The obtaining of adversarial examples is more challenging,
since the topological invariants capture the global properties
of the systems and are insensitive to local perturbations. We consider a three-band model for 3D
chiral topological insulators (TIs) \cite{neupert2012noncommutative,wang2014probe}: $H_{\text{TI}}  =  \sum_{\boldsymbol{k\in\text{BZ}}}\Psi_{\boldsymbol{k}}^{\dagger}H_{\boldsymbol{k}}\Psi_{\boldsymbol{k}}$,
where $\Psi_{\boldsymbol{k}}^{\dagger}=(c_{\boldsymbol{k},1}^{\dagger},c_{\boldsymbol{k},0}^{\dagger},c_{\boldsymbol{k},-1}^{\dagger})$
with $c_{\boldsymbol{k,\mu}}^{\dagger}$ the fermion creation operator
at momentum $\boldsymbol{k}=(k_{x},k_{y},k_{z})$ in the orbital (spin)
state $\mu=-1,0,1$ and the summation is over the Brillouin zone (BZ);
$H_{\boldsymbol{k}}=\lambda_{1}\sin k_{x}+\lambda_{2}\sin k_{y}+\lambda_{6}\sin k_{z}-\lambda_{7}(\cos k_{x}+\cos k_{y}+\cos k_{z}+h)$
denotes the single-particle Hamiltonian, with $\lambda_{1,2,6,7}$
being four traceless Gell-Mann matrices \cite{neupert2012noncommutative}. 
The topological properties for each band can
be characterized by a topological invariant $\chi^{(\eta)}$ \cite{TopInv} and it is straightforward to obtain that
$\chi^{(m)}=0,1,\text{and }-2$ for $|h|>3,1<|h|<3$, and $|h|<1$,
respectively. Recently, an experiment has been carried out to simulate $H_{\text{TI}}$
with the electron spins in a NV center
and a demonstration of the supervised learning approach to topological
phases has been reported \cite{Lian2019Machine}. Using
the measured density matrices in the momentum space (which
can be obtained through quantum state tomography) as input
data, a trained 3D convolutional neural network (CNN) can
correctly identify distinct topological phases with exceptionally high success probability, even when a large portion of the
experimentally generated raw data was dropped out or inaccessible. Here, we show that this approach is highly vulnerable to adversarial perturbations.

\begin{figure}[ht]
\includegraphics[width=0.45\textwidth]{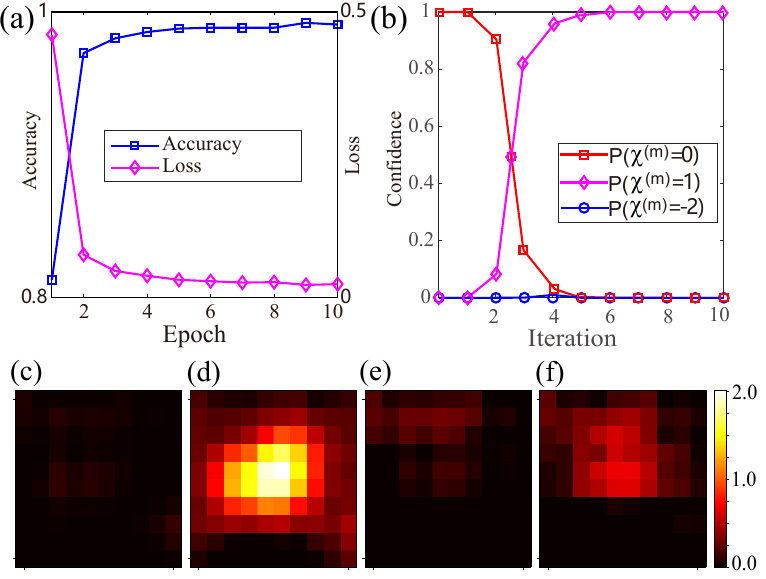}

\caption{
(a) The average accuracy and loss of the 3D convolutional neural network to classify the topological phases. (b) We use the momentum iterative method to obtain the adversarial examples. This
plot shows the classification probabilities as a function of the iteration number. After around two iterations, the network begin to misclassify the samples. \diff{(c-f) The activation maps (AM) of the sixth kernel in the first convolutional layer under different settings. (c) 
the average AM on all samples in the test set with $\chi^{(m)}=0$. (d) 
the average AM on $\chi^{(m)}=1$. (e) 
the AM obtained by taking the legitimate sample as the input to the topological phases classifier and (f) is taking the adversarial example as input.} \label{fig:3DTI}}
\end{figure}

\diff{We first train a 3D CNN with numerically simulated data. The training curve is shown in Fig. \ref{fig:3DTI}(a) and the accuracy on validation data saturates at a high value ($\approx99\%$) \cite{VMLPMSup}.} After
the training, we fix the model parameters of the CNN and utilize the
Fast Gradient Sign Method (FGSM) \cite{Madry2017Towards}, projected gradient descent (PGD) \cite{Madry2017Towards} and Momentum Iterative Method (MIM) \cite{dong2018boosting} to generate adversarial perturbations \cite{papernot2018cleverhans}. Fig. \ref{fig:3DTI}(b) shows the confidence
probabilities of the classification of a sample with $\chi^{({m})}=0$
as functions of the MIM iterations. From this figure, $P(\chi^{({m})}=0)$
decreases rapidly as the iteration number increases and converges
to a small value ($\approx2\%)$ after about eight iterations. Meanwhile,
$P(\chi^{({m})}=1)$ increases rapidly and converges to a large
value ($\approx98\%$), indicating a misclassification of the classifier\textemdash after
about eight iterations, the sample originally from the category $\chi^{({m})}=0$
is misclassified to belong to the category $\chi^{({m})}=1$
with a confidence level $\approx98\%$ \cite{VMLPMSup}. \diff{ We note that a direct calculation of the topological invariant through  integration confirms that $\chi^{(\text{m})}=0$ for this adversarial example, indicating that the tiny perturbation would not affect the traditional methods.}

\diff{It is more \diff{challenging }
to figure out why the topological phases classifier is so vulnerable to the adversarial perturbation. \diff{Since these convolutional kernels are repeatedly applied to different spatial windows, we have limited method to calculate the importance of each location, as we do with an Ising classifier.}
We study the activation maps of all convolutional kernels in the first convolutional layer. We find that the sixth kernel has totally different activation patterns for topologically trivial and nontrivial phases, which acts as a strong indicator for the classifier to distinguish these phases \cite{VMLPMSup}. Specifically, the activation patterns for $\chi^{(m)}=0,1$ phases are 
\diff{illustrated} in Fig.~\ref{fig:3DTI}(c-d).  We then compare the sixth kernel's activation maps of the legitimate sample and the adversarial example. As shown in Fig.~\ref{fig:3DTI}(e-f), we can clearly see that the tiny adversarial perturbation makes the activation map much more correlated with the $\chi^{(m)}=1$ ones, which gives the classifier high confidence to assert that the adversarial example belongs to the $\chi^{(m)}=1$ phase \cite{VMLPMSup}. This explains why adversarial examples can deceive the classifier.}

The above two examples clearly demonstrate the vulnerability of machine
learning approaches to classify different phases of matter. We mention
that, although we have only focused on these two examples, the existence
of adversarial perturbations is ubiquitous in learning various phases
(independent of the learning model and input data type) and the methods
used in the above examples can also be used to generate the desired
adversarial perturbations for different phases. From a more theoretical
computer science perspective, the vulnerability of the phase classifiers
can be understood as a consequence of the strong ``No Free Lunch''
theorem\textemdash there exists an intrinsic tension between adversarial
robustness and generalization accuracy \cite{tsipras2018robustness,fawzi2018adversarial,gilmer2018adversarial}.
The data distributions in the scenarios of learning phases of matter
typically satisfy the  $W_{2}$ Talagrand transportation-cost
inequality, thus a phase classifier could be adversarially deceived
with high probability \cite{dohmatob2018limitations}.

\begin{figure}[ht]
\includegraphics[width=0.49\textwidth]{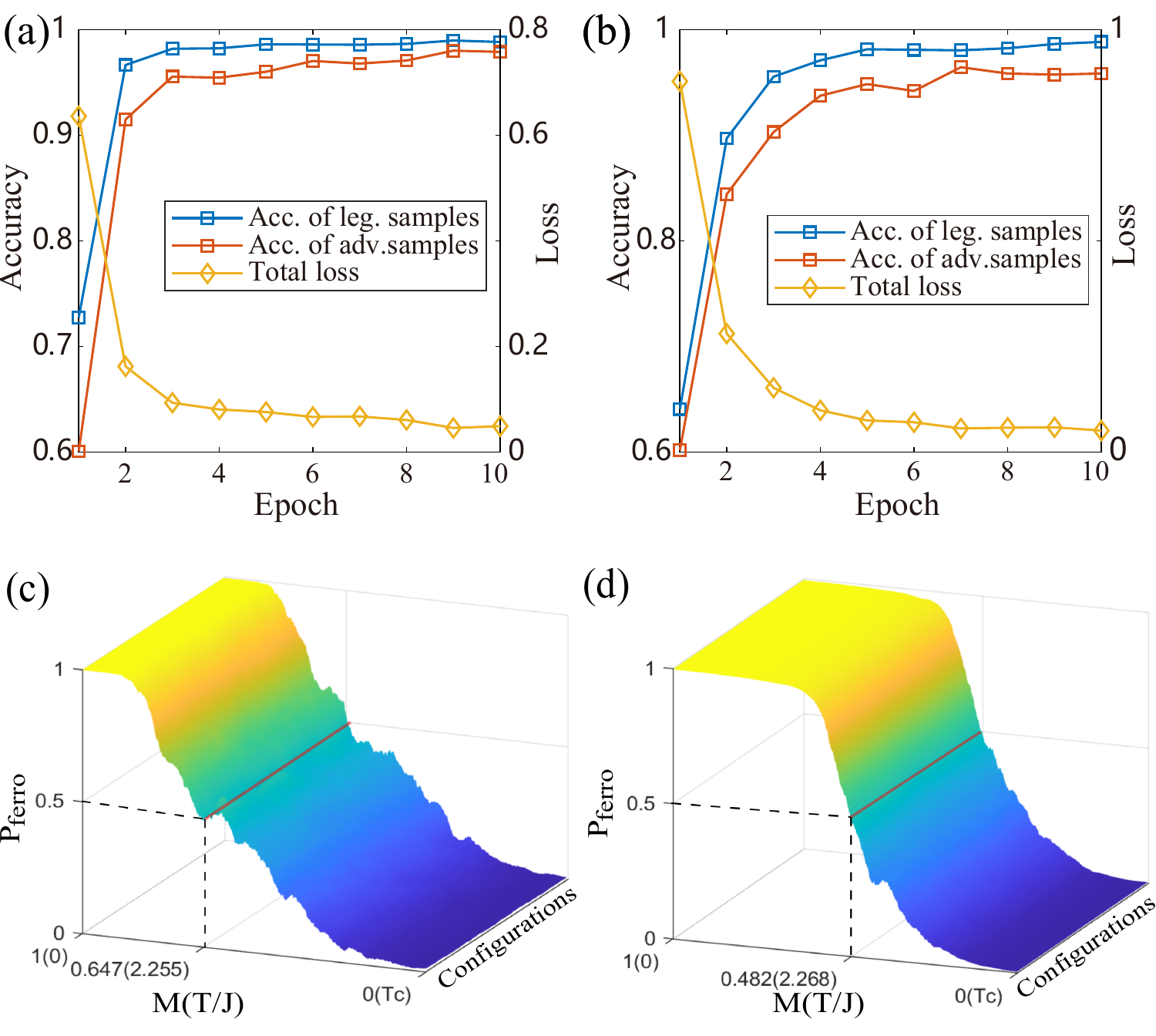}
\caption{The effectiveness of the adversarial training. (a) We first numerically generate adequate adversarial
examples with the FGSM, and then retrain the CNN with both the
legitimate and adversarial data. (b) Similar adversarial training for the defense of the PGD attack. \diff{(c) The original classifier's output representing the confidence of being ferromagnetic. The ``Configurations'' means different spin configurations with the same magnetization $M$. The corresponding temperature (listed in parentheses) of each $M$ is calculated by Onsager's formula \cite{Onsager1944Crystal}. The original classifier identify the transition point at $T=2.255$. (d) The refined classifier's output after the adversarial training against BIM attack. The identified transition point changes to $T=2.268$, which becomes closer to $T_c=2.269$.}  \label{fig:Adversarial-Training}}
\end{figure}

\textit{Adversarial training.}\textemdash In  adversarial
machine learning, a number of countermeasures against adversarial
examples have been developed \cite{chakraborty2018adversarial,yuan2019adversarial}.
\diff{Adversarial training, whose essential idea is to first generate a substantial amount of adversarial examples and then retrain the classifier with both original
data and crafted data, is one of these countermeasures to make the classifiers more robust. 
Here, in order to study how it works for machine learning phases of matter, we apply adversarial training to the 3D CNN classifier used in classifying topological phases.
Partial results are plotted in Fig. \ref{fig:Adversarial-Training}(a-b). While the classifier's performance on legitimate examples maintains intact,  the test accuracy on adversarial examples increases significantly (at the end of the adversarial training, it increases from about 60\% to 98\%). This result indicates that the retrained classifier is immune to the adversarial examples generated by the corresponding attacks. 
}

\diff{We also study how adversarial training can make the classifiers grasp physical quantities more thoroughly. As shown in Fig. \ref{fig:IsingM}(d), the activation map of the Ising model classifier becomes much flatter after adversarial training (the standard deviation of different positions is reduced from 0.88 to 0.20), which indicates that after adversarial learning the output of the classifier is more consistent with the physical order parameter of magnetization where each spin contributes equally, and hence more robust to adversarial perturbations. This is also reflected by the fact that after adversarial training the classifier can identify the phase transition point more accurately, as shown in Fig. \ref{fig:Adversarial-Training}(c-d).} 

\textit{Discussion and conclusion.\textemdash } \diff{ We mention that the adversarial training method is useful only on adversarial examples which are crafted
 on the original classifier. The defense may not work for black-box attacks \cite{narodytska2017simple,papernot2016transferability},
 where an adversary generates malicious examples on a locally trained
 substitute model. To deal with the transferred black-box attack, one
 may explore the recently proposed ensemble adversarial training method
 that retrain the classifier with adversarial examples generated from
 multiple sources \cite{tramer2017ensemble}. In the future, it would be interesting and desirable to find other defense strategies to strengthen the robustness of phase classifiers to adversarial perturbations.  In addition, an experimental demonstration of adversarial learning phases of matter together with defense strategies would also be an important step towards reliable practical applications of machine learning in physics. }

In summary, we have studied the robustness of machine learning approaches
in classifying different phases of matter. Our discussion is mainly
focused on supervised learning based on deep neural networks, but
its generalization to other types of learning models (such as unsupervised
learning or reinforcement learning) and other type of phases are
possible. Through two concrete examples, we have
demonstrated explicitly that typical phase classifiers based on deep
neural networks are extremely vulnerable to tiny adversarial perturbations. \diff{We have studied the explainability of adversarial examples and demonstrated that adversarial training significantly improves the robustness of phase classifiers by assisting the model to learn underlying physical principle and symmetries.}  
Our results reveal a novel vulnerability aspect for the growing field of machine learning phases of matter, which would benefit future studies
across condensed matter physics, machine learning, and artificial
intelligence.

We thank Christopher Monroe, John Preskill, Nana Liu, Peter Wittek, Ignacio Cirac, Roger Colbeck, Yi Zhang, Peter Zoller, Xiaopeng Li, Mucong Ding, Rainer Blatt, Zico Kolter, Juan Carrasquilla, and Peter Shor for helpful discussions. This work
was supported by the start-up fund from Tsinghua University (Grant
No. 53330300319). 

\bibliographystyle{apsrev4-1-title}
\bibliography{VMLPM-arVixNotes}

\begin{thebibliography}{79}%
\makeatletter
\providecommand \@ifxundefined [1]{%
 \@ifx{#1\undefined}
}%
\providecommand \@ifnum [1]{%
 \ifnum #1\expandafter \@firstoftwo
 \else \expandafter \@secondoftwo
 \fi
}%
\providecommand \@ifx [1]{%
 \ifx #1\expandafter \@firstoftwo
 \else \expandafter \@secondoftwo
 \fi
}%
\providecommand \natexlab [1]{#1}%
\providecommand \enquote  [1]{``#1''}%
\providecommand \bibnamefont  [1]{#1}%
\providecommand \bibfnamefont [1]{#1}%
\providecommand \citenamefont [1]{#1}%
\providecommand \href@noop [0]{\@secondoftwo}%
\providecommand \href [0]{\begingroup \@sanitize@url \@href}%
\providecommand \@href[1]{\@@startlink{#1}\@@href}%
\providecommand \@@href[1]{\endgroup#1\@@endlink}%
\providecommand \@sanitize@url [0]{\catcode `\\12\catcode `\$12\catcode
  `\&12\catcode `\#12\catcode `\^12\catcode `\_12\catcode `\%12\relax}%
\providecommand \@@startlink[1]{}%
\providecommand \@@endlink[0]{}%
\providecommand \url  [0]{\begingroup\@sanitize@url \@url }%
\providecommand \@url [1]{\endgroup\@href {#1}{\urlprefix }}%
\providecommand \urlprefix  [0]{URL }%
\providecommand \Eprint [0]{\href }%
\providecommand \doibase [0]{http://dx.doi.org/}%
\providecommand \selectlanguage [0]{\@gobble}%
\providecommand \bibinfo  [0]{\@secondoftwo}%
\providecommand \bibfield  [0]{\@secondoftwo}%
\providecommand \translation [1]{[#1]}%
\providecommand \BibitemOpen [0]{}%
\providecommand \bibitemStop [0]{}%
\providecommand \bibitemNoStop [0]{.\EOS\space}%
\providecommand \EOS [0]{\spacefactor3000\relax}%
\providecommand \BibitemShut  [1]{\csname bibitem#1\endcsname}%
\let\auto@bib@innerbib\@empty
\bibitem [{\citenamefont {LeCun}\ \emph {et~al.}(2015)\citenamefont {LeCun},
  \citenamefont {Bengio},\ and\ \citenamefont {Hinton}}]{Lecun2015Deep}%
  \BibitemOpen
  \bibfield  {author} {\bibinfo {author} {\bibfnamefont {Y.}~\bibnamefont
  {LeCun}}, \bibinfo {author} {\bibfnamefont {Y.}~\bibnamefont {Bengio}}, \
  and\ \bibinfo {author} {\bibfnamefont {G.}~\bibnamefont {Hinton}},\
  }\bibfield  {title} {\enquote {\bibinfo {title} {Deep learning},}\ }\href
  {\doibase 10.1038/nature14539} {\bibfield  {journal} {\bibinfo  {journal}
  {Nature}\ }\textbf {\bibinfo {volume} {521}},\ \bibinfo {pages} {436}
  (\bibinfo {year} {2015})}\BibitemShut {NoStop}%
\bibitem [{\citenamefont {Jordan}\ and\ \citenamefont
  {Mitchell}(2015)}]{Jordan2015Machine}%
  \BibitemOpen
  \bibfield  {author} {\bibinfo {author} {\bibfnamefont {M.}~\bibnamefont
  {Jordan}}\ and\ \bibinfo {author} {\bibfnamefont {T.}~\bibnamefont
  {Mitchell}},\ }\bibfield  {title} {\enquote {\bibinfo {title} {Machine
  learning: Trends, perspectives, and prospects},}\ }\href {\doibase
  10.1126/science.aaa8415} {\bibfield  {journal} {\bibinfo  {journal}
  {Science}\ }\textbf {\bibinfo {volume} {349}},\ \bibinfo {pages} {255}
  (\bibinfo {year} {2015})}\BibitemShut {NoStop}%
\bibitem [{\citenamefont {Sarma}\ \emph {et~al.}(2019)\citenamefont {Sarma},
  \citenamefont {Deng},\ and\ \citenamefont {Duan}}]{Sarma2019Machine}%
  \BibitemOpen
  \bibfield  {author} {\bibinfo {author} {\bibfnamefont {S.~D.}\ \bibnamefont
  {Sarma}}, \bibinfo {author} {\bibfnamefont {D.-L.}\ \bibnamefont {Deng}}, \
  and\ \bibinfo {author} {\bibfnamefont {L.-M.}\ \bibnamefont {Duan}},\
  }\bibfield  {title} {\enquote {\bibinfo {title} {Machine learning meets
  quantum physics},}\ }\href {\doibase 10.1063/PT.3.4164} {\bibfield  {journal}
  {\bibinfo  {journal} {Physics Today}\ }\textbf {\bibinfo {volume} {72}},\
  \bibinfo {pages} {48} (\bibinfo {year} {2019})}\BibitemShut {NoStop}%
\bibitem [{\citenamefont {Carleo}\ and\ \citenamefont
  {Troyer}(2017)}]{Carleo2016Solving}%
  \BibitemOpen
  \bibfield  {author} {\bibinfo {author} {\bibfnamefont {G.}~\bibnamefont
  {Carleo}}\ and\ \bibinfo {author} {\bibfnamefont {M.}~\bibnamefont
  {Troyer}},\ }\bibfield  {title} {\enquote {\bibinfo {title} {Solving the
  quantum many-body problem with artificial neural networks},}\ }\href
  {\doibase 10.1126/science.aag2302} {\bibfield  {journal} {\bibinfo  {journal}
  {Science}\ }\textbf {\bibinfo {volume} {355}},\ \bibinfo {pages} {602}
  (\bibinfo {year} {2017})}\BibitemShut {NoStop}%
\bibitem [{\citenamefont {Torlai}\ \emph {et~al.}(2018)\citenamefont {Torlai},
  \citenamefont {Mazzola}, \citenamefont {Carrasquilla}, \citenamefont
  {Troyer}, \citenamefont {Melko},\ and\ \citenamefont
  {Carleo}}]{Torlai2018Neural}%
  \BibitemOpen
  \bibfield  {author} {\bibinfo {author} {\bibfnamefont {G.}~\bibnamefont
  {Torlai}}, \bibinfo {author} {\bibfnamefont {G.}~\bibnamefont {Mazzola}},
  \bibinfo {author} {\bibfnamefont {J.}~\bibnamefont {Carrasquilla}}, \bibinfo
  {author} {\bibfnamefont {M.}~\bibnamefont {Troyer}}, \bibinfo {author}
  {\bibfnamefont {R.}~\bibnamefont {Melko}}, \ and\ \bibinfo {author}
  {\bibfnamefont {G.}~\bibnamefont {Carleo}},\ }\bibfield  {title} {\enquote
  {\bibinfo {title} {Neural-network quantum state tomography},}\ }\href
  {\doibase 10.1038/s41567-018-0048-5} {\bibfield  {journal} {\bibinfo
  {journal} {Nat. Phys.}\ }\textbf {\bibinfo {volume} {14}},\ \bibinfo {pages}
  {447} (\bibinfo {year} {2018})}\BibitemShut {NoStop}%
\bibitem [{\citenamefont {Nomura}\ \emph {et~al.}(2017)\citenamefont {Nomura},
  \citenamefont {Darmawan}, \citenamefont {Yamaji},\ and\ \citenamefont
  {Imada}}]{Nomura2017Restricted}%
  \BibitemOpen
  \bibfield  {author} {\bibinfo {author} {\bibfnamefont {Y.}~\bibnamefont
  {Nomura}}, \bibinfo {author} {\bibfnamefont {A.~S.}\ \bibnamefont
  {Darmawan}}, \bibinfo {author} {\bibfnamefont {Y.}~\bibnamefont {Yamaji}}, \
  and\ \bibinfo {author} {\bibfnamefont {M.}~\bibnamefont {Imada}},\ }\bibfield
   {title} {\enquote {\bibinfo {title} {Restricted boltzmann machine learning
  for solving strongly correlated quantum systems},}\ }\href {\doibase
  10.1103/PhysRevB.96.205152} {\bibfield  {journal} {\bibinfo  {journal} {Phys.
  Rev. B}\ }\textbf {\bibinfo {volume} {96}},\ \bibinfo {pages} {205152}
  (\bibinfo {year} {2017})}\BibitemShut {NoStop}%
\bibitem [{\citenamefont {You}\ \emph {et~al.}(2018)\citenamefont {You},
  \citenamefont {Yang},\ and\ \citenamefont {Qi}}]{You2017Machine}%
  \BibitemOpen
  \bibfield  {author} {\bibinfo {author} {\bibfnamefont {Y.-Z.}\ \bibnamefont
  {You}}, \bibinfo {author} {\bibfnamefont {Z.}~\bibnamefont {Yang}}, \ and\
  \bibinfo {author} {\bibfnamefont {X.-L.}\ \bibnamefont {Qi}},\ }\bibfield
  {title} {\enquote {\bibinfo {title} {Machine learning spatial geometry from
  entanglement features},}\ }\href {\doibase 10.1103/PhysRevB.97.045153}
  {\bibfield  {journal} {\bibinfo  {journal} {Phys. Rev. B}\ }\textbf {\bibinfo
  {volume} {97}},\ \bibinfo {pages} {045153} (\bibinfo {year}
  {2018})}\BibitemShut {NoStop}%
\bibitem [{\citenamefont {Deng}\ \emph
  {et~al.}(2017{\natexlab{a}})\citenamefont {Deng}, \citenamefont {Li},\ and\
  \citenamefont {Das~Sarma}}]{Deng2017Machine}%
  \BibitemOpen
  \bibfield  {author} {\bibinfo {author} {\bibfnamefont {D.-L.}\ \bibnamefont
  {Deng}}, \bibinfo {author} {\bibfnamefont {X.}~\bibnamefont {Li}}, \ and\
  \bibinfo {author} {\bibfnamefont {S.}~\bibnamefont {Das~Sarma}},\ }\bibfield
  {title} {\enquote {\bibinfo {title} {Machine learning topological states},}\
  }\href {\doibase 10.1103/PhysRevB.96.195145} {\bibfield  {journal} {\bibinfo
  {journal} {Phys. Rev. B}\ }\textbf {\bibinfo {volume} {96}},\ \bibinfo
  {pages} {195145} (\bibinfo {year} {2017}{\natexlab{a}})}\BibitemShut
  {NoStop}%
\bibitem [{\citenamefont {Deng}(2018)}]{Deng2017MachineBN}%
  \BibitemOpen
  \bibfield  {author} {\bibinfo {author} {\bibfnamefont {D.-L.}\ \bibnamefont
  {Deng}},\ }\bibfield  {title} {\enquote {\bibinfo {title} {Machine learning
  detection of bell nonlocality in quantum many-body systems},}\ }\href
  {\doibase 10.1103/PhysRevLett.120.240402} {\bibfield  {journal} {\bibinfo
  {journal} {Phys. Rev. Lett.}\ }\textbf {\bibinfo {volume} {120}},\ \bibinfo
  {pages} {240402} (\bibinfo {year} {2018})}\BibitemShut {NoStop}%
\bibitem [{\citenamefont {Deng}\ \emph
  {et~al.}(2017{\natexlab{b}})\citenamefont {Deng}, \citenamefont {Li},\ and\
  \citenamefont {Das~Sarma}}]{Deng2017Quantum}%
  \BibitemOpen
  \bibfield  {author} {\bibinfo {author} {\bibfnamefont {D.-L.}\ \bibnamefont
  {Deng}}, \bibinfo {author} {\bibfnamefont {X.}~\bibnamefont {Li}}, \ and\
  \bibinfo {author} {\bibfnamefont {S.}~\bibnamefont {Das~Sarma}},\ }\bibfield
  {title} {\enquote {\bibinfo {title} {Quantum entanglement in neural network
  states},}\ }\href {\doibase 10.1103/PhysRevX.7.021021} {\bibfield  {journal}
  {\bibinfo  {journal} {Phys. Rev. X}\ }\textbf {\bibinfo {volume} {7}},\
  \bibinfo {pages} {021021} (\bibinfo {year} {2017}{\natexlab{b}})}\BibitemShut
  {NoStop}%
\bibitem [{\citenamefont {Gao}\ and\ \citenamefont
  {Duan}(2017)}]{Gao2017Efficient}%
  \BibitemOpen
  \bibfield  {author} {\bibinfo {author} {\bibfnamefont {X.}~\bibnamefont
  {Gao}}\ and\ \bibinfo {author} {\bibfnamefont {L.-M.}\ \bibnamefont {Duan}},\
  }\bibfield  {title} {\enquote {\bibinfo {title} {Efficient representation of
  quantum many-body states with deep neural networks},}\ }\href {\doibase
  10.1038/s41467-017-00705-2} {\bibfield  {journal} {\bibinfo  {journal} {Nat.
  Commun.}\ }\textbf {\bibinfo {volume} {8}},\ \bibinfo {pages} {662} (\bibinfo
  {year} {2017})}\BibitemShut {NoStop}%
\bibitem [{\citenamefont {Melko}\ \emph {et~al.}(2019)\citenamefont {Melko},
  \citenamefont {Carleo}, \citenamefont {Carrasquilla},\ and\ \citenamefont
  {Cirac}}]{melko2019restricted}%
  \BibitemOpen
  \bibfield  {author} {\bibinfo {author} {\bibfnamefont {R.~G.}\ \bibnamefont
  {Melko}}, \bibinfo {author} {\bibfnamefont {G.}~\bibnamefont {Carleo}},
  \bibinfo {author} {\bibfnamefont {J.}~\bibnamefont {Carrasquilla}}, \ and\
  \bibinfo {author} {\bibfnamefont {J.~I.}\ \bibnamefont {Cirac}},\ }\bibfield
  {title} {\enquote {\bibinfo {title} {Restricted {{Boltzmann}} machines in
  quantum physics},}\ }\href {\doibase 10.1038/s41567-019-0545-1} {\bibfield
  {journal} {\bibinfo  {journal} {Nat. Phys.}\ }\textbf {\bibinfo {volume}
  {15}},\ \bibinfo {pages} {887} (\bibinfo {year} {2019})}\BibitemShut
  {NoStop}%
\bibitem [{\citenamefont {Ch'ng}\ \emph {et~al.}(2017)\citenamefont {Ch'ng},
  \citenamefont {Carrasquilla}, \citenamefont {Melko},\ and\ \citenamefont
  {Khatami}}]{Chng2017Machine}%
  \BibitemOpen
  \bibfield  {author} {\bibinfo {author} {\bibfnamefont {K.}~\bibnamefont
  {Ch'ng}}, \bibinfo {author} {\bibfnamefont {J.}~\bibnamefont {Carrasquilla}},
  \bibinfo {author} {\bibfnamefont {R.~G.}\ \bibnamefont {Melko}}, \ and\
  \bibinfo {author} {\bibfnamefont {E.}~\bibnamefont {Khatami}},\ }\bibfield
  {title} {\enquote {\bibinfo {title} {Machine learning phases of strongly
  correlated fermions},}\ }\href {\doibase 10.1103/PhysRevX.7.031038}
  {\bibfield  {journal} {\bibinfo  {journal} {Phys. Rev. X}\ }\textbf {\bibinfo
  {volume} {7}},\ \bibinfo {pages} {031038} (\bibinfo {year}
  {2017})}\BibitemShut {NoStop}%
\bibitem [{\citenamefont {Wang}(2016)}]{Wang2016Discovering}%
  \BibitemOpen
  \bibfield  {author} {\bibinfo {author} {\bibfnamefont {L.}~\bibnamefont
  {Wang}},\ }\bibfield  {title} {\enquote {\bibinfo {title} {Discovering phase
  transitions with unsupervised learning},}\ }\href {\doibase
  10.1103/PhysRevB.94.195105} {\bibfield  {journal} {\bibinfo  {journal} {Phys.
  Rev. B}\ }\textbf {\bibinfo {volume} {94}},\ \bibinfo {pages} {195105}
  (\bibinfo {year} {2016})}\BibitemShut {NoStop}%
\bibitem [{\citenamefont {Zhang}\ and\ \citenamefont
  {Kim}(2017)}]{Zhang2016Triangular}%
  \BibitemOpen
  \bibfield  {author} {\bibinfo {author} {\bibfnamefont {Y.}~\bibnamefont
  {Zhang}}\ and\ \bibinfo {author} {\bibfnamefont {E.-A.}\ \bibnamefont
  {Kim}},\ }\bibfield  {title} {\enquote {\bibinfo {title} {Quantum loop
  topography for machine learning},}\ }\href {\doibase
  10.1103/PhysRevLett.118.216401} {\bibfield  {journal} {\bibinfo  {journal}
  {Phys. Rev. Lett.}\ }\textbf {\bibinfo {volume} {118}},\ \bibinfo {pages}
  {216401} (\bibinfo {year} {2017})}\BibitemShut {NoStop}%
\bibitem [{\citenamefont {Carrasquilla}\ and\ \citenamefont
  {Melko}(2017)}]{Carrasquilla2017Machine}%
  \BibitemOpen
  \bibfield  {author} {\bibinfo {author} {\bibfnamefont {J.}~\bibnamefont
  {Carrasquilla}}\ and\ \bibinfo {author} {\bibfnamefont {R.~G.}\ \bibnamefont
  {Melko}},\ }\bibfield  {title} {\enquote {\bibinfo {title} {Machine learning
  phases of matter},}\ }\href {\doibase 10.1038/nphys4035} {\bibfield
  {journal} {\bibinfo  {journal} {Nat. Phys.}\ }\textbf {\bibinfo {volume}
  {13}},\ \bibinfo {pages} {431} (\bibinfo {year} {2017})}\BibitemShut
  {NoStop}%
\bibitem [{\citenamefont {van Nieuwenburg}\ \emph {et~al.}(2017)\citenamefont
  {van Nieuwenburg}, \citenamefont {Liu},\ and\ \citenamefont
  {Huber}}]{van2017Learning}%
  \BibitemOpen
  \bibfield  {author} {\bibinfo {author} {\bibfnamefont {E.~P.}\ \bibnamefont
  {van Nieuwenburg}}, \bibinfo {author} {\bibfnamefont {Y.-H.}\ \bibnamefont
  {Liu}}, \ and\ \bibinfo {author} {\bibfnamefont {S.~D.}\ \bibnamefont
  {Huber}},\ }\bibfield  {title} {\enquote {\bibinfo {title} {Learning phase
  transitions by confusion},}\ }\href {\doibase 10.1038/nphys4037} {\bibfield
  {journal} {\bibinfo  {journal} {Nat. Phys.}\ }\textbf {\bibinfo {volume}
  {13}},\ \bibinfo {pages} {435} (\bibinfo {year} {2017})}\BibitemShut
  {NoStop}%
\bibitem [{\citenamefont {Broecker}\ \emph {et~al.}(2017)\citenamefont
  {Broecker}, \citenamefont {Carrasquilla}, \citenamefont {Melko},\ and\
  \citenamefont {Trebst}}]{Broecker2017Machine}%
  \BibitemOpen
  \bibfield  {author} {\bibinfo {author} {\bibfnamefont {P.}~\bibnamefont
  {Broecker}}, \bibinfo {author} {\bibfnamefont {J.}~\bibnamefont
  {Carrasquilla}}, \bibinfo {author} {\bibfnamefont {R.~G.}\ \bibnamefont
  {Melko}}, \ and\ \bibinfo {author} {\bibfnamefont {S.}~\bibnamefont
  {Trebst}},\ }\bibfield  {title} {\enquote {\bibinfo {title} {Machine learning
  quantum phases of matter beyond the fermion sign problem},}\ }\href {\doibase
  10.1038/s41598-017-09098-0} {\bibfield  {journal} {\bibinfo  {journal} {Sci.
  Rep.}\ }\textbf {\bibinfo {volume} {7}},\ \bibinfo {pages} {8823} (\bibinfo
  {year} {2017})}\BibitemShut {NoStop}%
\bibitem [{\citenamefont {Wetzel}(2017)}]{Wetzel2017Unsupervised}%
  \BibitemOpen
  \bibfield  {author} {\bibinfo {author} {\bibfnamefont {S.~J.}\ \bibnamefont
  {Wetzel}},\ }\bibfield  {title} {\enquote {\bibinfo {title} {Unsupervised
  learning of phase transitions: From principal component analysis to
  variational autoencoders},}\ }\href {\doibase 10.1103/PhysRevE.96.022140}
  {\bibfield  {journal} {\bibinfo  {journal} {Phys. Rev. E}\ }\textbf {\bibinfo
  {volume} {96}},\ \bibinfo {pages} {022140} (\bibinfo {year}
  {2017})}\BibitemShut {NoStop}%
\bibitem [{\citenamefont {Hu}\ \emph {et~al.}(2017)\citenamefont {Hu},
  \citenamefont {Singh},\ and\ \citenamefont {Scalettar}}]{Hu2017Discovering}%
  \BibitemOpen
  \bibfield  {author} {\bibinfo {author} {\bibfnamefont {W.}~\bibnamefont
  {Hu}}, \bibinfo {author} {\bibfnamefont {R.~R.~P.}\ \bibnamefont {Singh}}, \
  and\ \bibinfo {author} {\bibfnamefont {R.~T.}\ \bibnamefont {Scalettar}},\
  }\bibfield  {title} {\enquote {\bibinfo {title} {Discovering phases, phase
  transitions, and crossovers through unsupervised machine learning: A critical
  examination},}\ }\href {\doibase 10.1103/PhysRevE.95.062122} {\bibfield
  {journal} {\bibinfo  {journal} {Phys. Rev. E}\ }\textbf {\bibinfo {volume}
  {95}},\ \bibinfo {pages} {062122} (\bibinfo {year} {2017})}\BibitemShut
  {NoStop}%
\bibitem [{\citenamefont {Hsu}\ \emph {et~al.}(2018)\citenamefont {Hsu},
  \citenamefont {Li}, \citenamefont {Deng},\ and\ \citenamefont
  {Das~Sarma}}]{Hsu2018Machine}%
  \BibitemOpen
  \bibfield  {author} {\bibinfo {author} {\bibfnamefont {Y.-T.}\ \bibnamefont
  {Hsu}}, \bibinfo {author} {\bibfnamefont {X.}~\bibnamefont {Li}}, \bibinfo
  {author} {\bibfnamefont {D.-L.}\ \bibnamefont {Deng}}, \ and\ \bibinfo
  {author} {\bibfnamefont {S.}~\bibnamefont {Das~Sarma}},\ }\bibfield  {title}
  {\enquote {\bibinfo {title} {Machine learning many-body localization: Search
  for the elusive nonergodic metal},}\ }\href {\doibase
  10.1103/PhysRevLett.121.245701} {\bibfield  {journal} {\bibinfo  {journal}
  {Phys. Rev. Lett.}\ }\textbf {\bibinfo {volume} {121}},\ \bibinfo {pages}
  {245701} (\bibinfo {year} {2018})}\BibitemShut {NoStop}%
\bibitem [{\citenamefont {Rodriguez-Nieva}\ and\ \citenamefont
  {Scheurer}(2019)}]{rodriguez2019identifying}%
  \BibitemOpen
  \bibfield  {author} {\bibinfo {author} {\bibfnamefont {J.~F.}\ \bibnamefont
  {Rodriguez-Nieva}}\ and\ \bibinfo {author} {\bibfnamefont {M.~S.}\
  \bibnamefont {Scheurer}},\ }\bibfield  {title} {\enquote {\bibinfo {title}
  {Identifying topological order through unsupervised machine learning},}\
  }\href {\doibase 10.1038/s41567-019-0512-x} {\bibfield  {journal} {\bibinfo
  {journal} {Nat. Phys.}\ ,\ \bibinfo {pages} {790}} (\bibinfo {year}
  {2019})}\BibitemShut {NoStop}%
\bibitem [{\citenamefont {Zhang}\ \emph {et~al.}(2018)\citenamefont {Zhang},
  \citenamefont {Shen},\ and\ \citenamefont {Zhai}}]{Zhang2018Machine}%
  \BibitemOpen
  \bibfield  {author} {\bibinfo {author} {\bibfnamefont {P.}~\bibnamefont
  {Zhang}}, \bibinfo {author} {\bibfnamefont {H.}~\bibnamefont {Shen}}, \ and\
  \bibinfo {author} {\bibfnamefont {H.}~\bibnamefont {Zhai}},\ }\bibfield
  {title} {\enquote {\bibinfo {title} {Machine learning topological invariants
  with neural networks},}\ }\href {\doibase 10.1103/PhysRevLett.120.066401}
  {\bibfield  {journal} {\bibinfo  {journal} {Phys. Rev. Lett.}\ }\textbf
  {\bibinfo {volume} {120}},\ \bibinfo {pages} {066401} (\bibinfo {year}
  {2018})}\BibitemShut {NoStop}%
\bibitem [{\citenamefont {Huembeli}\ \emph {et~al.}(2018)\citenamefont
  {Huembeli}, \citenamefont {Dauphin},\ and\ \citenamefont
  {Wittek}}]{Huembeli2018Identifying}%
  \BibitemOpen
  \bibfield  {author} {\bibinfo {author} {\bibfnamefont {P.}~\bibnamefont
  {Huembeli}}, \bibinfo {author} {\bibfnamefont {A.}~\bibnamefont {Dauphin}}, \
  and\ \bibinfo {author} {\bibfnamefont {P.}~\bibnamefont {Wittek}},\
  }\bibfield  {title} {\enquote {\bibinfo {title} {Identifying quantum phase
  transitions with adversarial neural networks},}\ }\href {\doibase
  10.1103/PhysRevB.97.134109} {\bibfield  {journal} {\bibinfo  {journal} {Phys.
  Rev. B}\ }\textbf {\bibinfo {volume} {97}},\ \bibinfo {pages} {134109}
  (\bibinfo {year} {2018})}\BibitemShut {NoStop}%
\bibitem [{\citenamefont {Suchsland}\ and\ \citenamefont
  {Wessel}(2018)}]{Suchsland2018Parameter}%
  \BibitemOpen
  \bibfield  {author} {\bibinfo {author} {\bibfnamefont {P.}~\bibnamefont
  {Suchsland}}\ and\ \bibinfo {author} {\bibfnamefont {S.}~\bibnamefont
  {Wessel}},\ }\bibfield  {title} {\enquote {\bibinfo {title} {Parameter
  diagnostics of phases and phase transition learning by neural networks},}\
  }\href {\doibase 10.1103/PhysRevB.97.174435} {\bibfield  {journal} {\bibinfo
  {journal} {Phys. Rev. B}\ }\textbf {\bibinfo {volume} {97}},\ \bibinfo
  {pages} {174435} (\bibinfo {year} {2018})}\BibitemShut {NoStop}%
\bibitem [{\citenamefont {Ohtsuki}\ and\ \citenamefont
  {Ohtsuki}(2016)}]{ohtsuki2016deep}%
  \BibitemOpen
  \bibfield  {author} {\bibinfo {author} {\bibfnamefont {T.}~\bibnamefont
  {Ohtsuki}}\ and\ \bibinfo {author} {\bibfnamefont {T.}~\bibnamefont
  {Ohtsuki}},\ }\bibfield  {title} {\enquote {\bibinfo {title} {Deep learning
  the quantum phase transitions in random two-dimensional electron systems},}\
  }\href {https://journals.jps.jp/doi/full/10.7566/JPSJ.85.123706} {\bibfield
  {journal} {\bibinfo  {journal} {J. Phys. Soc. Japan}\ }\textbf {\bibinfo
  {volume} {85}},\ \bibinfo {pages} {123706} (\bibinfo {year}
  {2016})}\BibitemShut {NoStop}%
\bibitem [{\citenamefont {Ohtsuki}\ and\ \citenamefont
  {Ohtsuki}(2017)}]{ohtsuki2017deep}%
  \BibitemOpen
  \bibfield  {author} {\bibinfo {author} {\bibfnamefont {T.}~\bibnamefont
  {Ohtsuki}}\ and\ \bibinfo {author} {\bibfnamefont {T.}~\bibnamefont
  {Ohtsuki}},\ }\bibfield  {title} {\enquote {\bibinfo {title} {Deep learning
  the quantum phase transitions in random electron systems: Applications to
  three dimensions},}\ }\href
  {https://journals.jps.jp/doi/full/10.7566/JPSJ.86.044708} {\bibfield
  {journal} {\bibinfo  {journal} {J. Phys. Soc. Japan}\ }\textbf {\bibinfo
  {volume} {86}},\ \bibinfo {pages} {044708} (\bibinfo {year}
  {2017})}\BibitemShut {NoStop}%
\bibitem [{\citenamefont {Ohtsuki}\ and\ \citenamefont
  {Mano}(2020)}]{ohtsuki2019drawing}%
  \BibitemOpen
  \bibfield  {author} {\bibinfo {author} {\bibfnamefont {T.}~\bibnamefont
  {Ohtsuki}}\ and\ \bibinfo {author} {\bibfnamefont {T.}~\bibnamefont {Mano}},\
  }\bibfield  {title} {\enquote {\bibinfo {title} {Drawing phase diagrams of
  random quantum systems by deep learning the wave functions},}\ }\href
  {\doibase 10.7566/jpsj.89.022001} {\bibfield  {journal} {\bibinfo  {journal}
  {J. Phys. Soc. Japan}\ }\textbf {\bibinfo {volume} {89}},\ \bibinfo {pages}
  {022001} (\bibinfo {year} {2020})}\BibitemShut {NoStop}%
\bibitem [{\citenamefont {Greplova}\ \emph {et~al.}(2020)\citenamefont
  {Greplova}, \citenamefont {Valenti}, \citenamefont {Boschung}, \citenamefont
  {Sch{\"a}fer}, \citenamefont {L{\"o}rch},\ and\ \citenamefont
  {Huber}}]{greplova2019unsupervised}%
  \BibitemOpen
  \bibfield  {author} {\bibinfo {author} {\bibfnamefont {E.}~\bibnamefont
  {Greplova}}, \bibinfo {author} {\bibfnamefont {A.}~\bibnamefont {Valenti}},
  \bibinfo {author} {\bibfnamefont {G.}~\bibnamefont {Boschung}}, \bibinfo
  {author} {\bibfnamefont {F.}~\bibnamefont {Sch{\"a}fer}}, \bibinfo {author}
  {\bibfnamefont {N.}~\bibnamefont {L{\"o}rch}}, \ and\ \bibinfo {author}
  {\bibfnamefont {S.}~\bibnamefont {Huber}},\ }\bibfield  {title} {\enquote
  {\bibinfo {title} {Unsupervised identification of topological order using
  predictive models},}\ }\href {\doibase 10.1088/1367-2630/ab7771} {\bibfield
  {journal} {\bibinfo  {journal} {New J. Phys.}\ }\textbf {\bibinfo {volume}
  {22}},\ \bibinfo {pages} {045003} (\bibinfo {year} {2020})}\BibitemShut
  {NoStop}%
\bibitem [{\citenamefont {Vargas-Hern\'andez}\ \emph
  {et~al.}(2018)\citenamefont {Vargas-Hern\'andez}, \citenamefont {Sous},
  \citenamefont {Berciu},\ and\ \citenamefont
  {Krems}}]{Hernandez2018Extrapolating}%
  \BibitemOpen
  \bibfield  {author} {\bibinfo {author} {\bibfnamefont {R.~A.}\ \bibnamefont
  {Vargas-Hern\'andez}}, \bibinfo {author} {\bibfnamefont {J.}~\bibnamefont
  {Sous}}, \bibinfo {author} {\bibfnamefont {M.}~\bibnamefont {Berciu}}, \ and\
  \bibinfo {author} {\bibfnamefont {R.~V.}\ \bibnamefont {Krems}},\ }\bibfield
  {title} {\enquote {\bibinfo {title} {Extrapolating quantum observables with
  machine learning: Inferring multiple phase transitions from properties of a
  single phase},}\ }\href {\doibase 10.1103/PhysRevLett.121.255702} {\bibfield
  {journal} {\bibinfo  {journal} {Phys. Rev. Lett.}\ }\textbf {\bibinfo
  {volume} {121}},\ \bibinfo {pages} {255702} (\bibinfo {year}
  {2018})}\BibitemShut {NoStop}%
\bibitem [{\citenamefont {Lian}\ \emph {et~al.}(2019)\citenamefont {Lian},
  \citenamefont {Wang}, \citenamefont {Lu}, \citenamefont {Huang},
  \citenamefont {Wang}, \citenamefont {Yuan}, \citenamefont {Zhang},
  \citenamefont {Ouyang}, \citenamefont {Wang}, \citenamefont {Huang},
  \citenamefont {He}, \citenamefont {Chang}, \citenamefont {Deng},\ and\
  \citenamefont {Duan}}]{Lian2019Machine}%
  \BibitemOpen
  \bibfield  {author} {\bibinfo {author} {\bibfnamefont {W.}~\bibnamefont
  {Lian}}, \bibinfo {author} {\bibfnamefont {S.-T.}\ \bibnamefont {Wang}},
  \bibinfo {author} {\bibfnamefont {S.}~\bibnamefont {Lu}}, \bibinfo {author}
  {\bibfnamefont {Y.}~\bibnamefont {Huang}}, \bibinfo {author} {\bibfnamefont
  {F.}~\bibnamefont {Wang}}, \bibinfo {author} {\bibfnamefont {X.}~\bibnamefont
  {Yuan}}, \bibinfo {author} {\bibfnamefont {W.}~\bibnamefont {Zhang}},
  \bibinfo {author} {\bibfnamefont {X.}~\bibnamefont {Ouyang}}, \bibinfo
  {author} {\bibfnamefont {X.}~\bibnamefont {Wang}}, \bibinfo {author}
  {\bibfnamefont {X.}~\bibnamefont {Huang}}, \bibinfo {author} {\bibfnamefont
  {L.}~\bibnamefont {He}}, \bibinfo {author} {\bibfnamefont {X.}~\bibnamefont
  {Chang}}, \bibinfo {author} {\bibfnamefont {D.-L.}\ \bibnamefont {Deng}}, \
  and\ \bibinfo {author} {\bibfnamefont {L.-M.}\ \bibnamefont {Duan}},\
  }\bibfield  {title} {\enquote {\bibinfo {title} {Machine learning topological
  phases with a solid-state quantum simulator},}\ }\href {\doibase
  10.1103/PhysRevLett.122.210503} {\bibfield  {journal} {\bibinfo  {journal}
  {Phys. Rev. Lett.}\ }\textbf {\bibinfo {volume} {122}},\ \bibinfo {pages}
  {210503} (\bibinfo {year} {2019})}\BibitemShut {NoStop}%
\bibitem [{\citenamefont {Rem}\ \emph {et~al.}(2019)\citenamefont {Rem},
  \citenamefont {K{\"a}ming}, \citenamefont {Tarnowski}, \citenamefont
  {Asteria}, \citenamefont {Fl{\"a}schner}, \citenamefont {Becker},
  \citenamefont {Sengstock},\ and\ \citenamefont
  {Weitenberg}}]{rem2019identifying}%
  \BibitemOpen
  \bibfield  {author} {\bibinfo {author} {\bibfnamefont {B.~S.}\ \bibnamefont
  {Rem}}, \bibinfo {author} {\bibfnamefont {N.}~\bibnamefont {K{\"a}ming}},
  \bibinfo {author} {\bibfnamefont {M.}~\bibnamefont {Tarnowski}}, \bibinfo
  {author} {\bibfnamefont {L.}~\bibnamefont {Asteria}}, \bibinfo {author}
  {\bibfnamefont {N.}~\bibnamefont {Fl{\"a}schner}}, \bibinfo {author}
  {\bibfnamefont {C.}~\bibnamefont {Becker}}, \bibinfo {author} {\bibfnamefont
  {K.}~\bibnamefont {Sengstock}}, \ and\ \bibinfo {author} {\bibfnamefont
  {C.}~\bibnamefont {Weitenberg}},\ }\bibfield  {title} {\enquote {\bibinfo
  {title} {Identifying quantum phase transitions using artificial neural
  networks on experimental data},}\ }\href {\doibase 10.1038/s41567-019-0554-0}
  {\bibfield  {journal} {\bibinfo  {journal} {Nat. Phys.}\ }\textbf {\bibinfo
  {volume} {15}},\ \bibinfo {pages} {917} (\bibinfo {year} {2019})}\BibitemShut
  {NoStop}%
\bibitem [{\citenamefont {Bohrdt}\ \emph {et~al.}(2019)\citenamefont {Bohrdt},
  \citenamefont {Chiu}, \citenamefont {Ji}, \citenamefont {Xu}, \citenamefont
  {Greif}, \citenamefont {Greiner}, \citenamefont {Demler}, \citenamefont
  {Grusdt},\ and\ \citenamefont {Knap}}]{bohrdt2019classifying}%
  \BibitemOpen
  \bibfield  {author} {\bibinfo {author} {\bibfnamefont {A.}~\bibnamefont
  {Bohrdt}}, \bibinfo {author} {\bibfnamefont {C.~S.}\ \bibnamefont {Chiu}},
  \bibinfo {author} {\bibfnamefont {G.}~\bibnamefont {Ji}}, \bibinfo {author}
  {\bibfnamefont {M.}~\bibnamefont {Xu}}, \bibinfo {author} {\bibfnamefont
  {D.}~\bibnamefont {Greif}}, \bibinfo {author} {\bibfnamefont
  {M.}~\bibnamefont {Greiner}}, \bibinfo {author} {\bibfnamefont
  {E.}~\bibnamefont {Demler}}, \bibinfo {author} {\bibfnamefont
  {F.}~\bibnamefont {Grusdt}}, \ and\ \bibinfo {author} {\bibfnamefont
  {M.}~\bibnamefont {Knap}},\ }\bibfield  {title} {\enquote {\bibinfo {title}
  {Classifying snapshots of the doped hubbard model with machine learning},}\
  }\href {\doibase 10.1038/s41567-019-0565-x} {\bibfield  {journal} {\bibinfo
  {journal} {Nat. Phys.}\ }\textbf {\bibinfo {volume} {15}},\ \bibinfo {pages}
  {921} (\bibinfo {year} {2019})}\BibitemShut {NoStop}%
\bibitem [{\citenamefont {Zhang}\ \emph {et~al.}(2019)\citenamefont {Zhang},
  \citenamefont {Mesaros}, \citenamefont {Fujita}, \citenamefont {Edkins},
  \citenamefont {Hamidian}, \citenamefont {Ch'ng}, \citenamefont {Eisaki},
  \citenamefont {Uchida}, \citenamefont {Davis}, \citenamefont {Khatami} \emph
  {et~al.}}]{zhang2019machine}%
  \BibitemOpen
  \bibfield  {author} {\bibinfo {author} {\bibfnamefont {Y.}~\bibnamefont
  {Zhang}}, \bibinfo {author} {\bibfnamefont {A.}~\bibnamefont {Mesaros}},
  \bibinfo {author} {\bibfnamefont {K.}~\bibnamefont {Fujita}}, \bibinfo
  {author} {\bibfnamefont {S.}~\bibnamefont {Edkins}}, \bibinfo {author}
  {\bibfnamefont {M.}~\bibnamefont {Hamidian}}, \bibinfo {author}
  {\bibfnamefont {K.}~\bibnamefont {Ch'ng}}, \bibinfo {author} {\bibfnamefont
  {H.}~\bibnamefont {Eisaki}}, \bibinfo {author} {\bibfnamefont
  {S.}~\bibnamefont {Uchida}}, \bibinfo {author} {\bibfnamefont {J.~S.}\
  \bibnamefont {Davis}}, \bibinfo {author} {\bibfnamefont {E.}~\bibnamefont
  {Khatami}},  \emph {et~al.},\ }\bibfield  {title} {\enquote {\bibinfo {title}
  {Machine learning in electronic-quantum-matter imaging experiments},}\ }\href
  {\doibase 10.1038/s41586-019-1319-8} {\bibfield  {journal} {\bibinfo
  {journal} {Nature}\ }\textbf {\bibinfo {volume} {570}},\ \bibinfo {pages}
  {484} (\bibinfo {year} {2019})}\BibitemShut {NoStop}%
\bibitem [{\citenamefont {Biggio}\ and\ \citenamefont
  {Roli}(2018)}]{biggio2018wild}%
  \BibitemOpen
  \bibfield  {author} {\bibinfo {author} {\bibfnamefont {B.}~\bibnamefont
  {Biggio}}\ and\ \bibinfo {author} {\bibfnamefont {F.}~\bibnamefont {Roli}},\
  }\bibfield  {title} {\enquote {\bibinfo {title} {Wild patterns: Ten years
  after the rise of adversarial machine learning},}\ }\href
  {https://doi.org/10.1016/j.patcog.2018.07.023} {\bibfield  {journal}
  {\bibinfo  {journal} {Pattern Recognition}\ }\textbf {\bibinfo {volume}
  {84}},\ \bibinfo {pages} {317} (\bibinfo {year} {2018})}\BibitemShut
  {NoStop}%
\bibitem [{\citenamefont {Huang}\ \emph {et~al.}(2011)\citenamefont {Huang},
  \citenamefont {Joseph}, \citenamefont {Nelson}, \citenamefont {Rubinstein},\
  and\ \citenamefont {Tygar}}]{Huang2011Adversarial}%
  \BibitemOpen
  \bibfield  {author} {\bibinfo {author} {\bibfnamefont {L.}~\bibnamefont
  {Huang}}, \bibinfo {author} {\bibfnamefont {A.~D.}\ \bibnamefont {Joseph}},
  \bibinfo {author} {\bibfnamefont {B.}~\bibnamefont {Nelson}}, \bibinfo
  {author} {\bibfnamefont {B.~I.}\ \bibnamefont {Rubinstein}}, \ and\ \bibinfo
  {author} {\bibfnamefont {J.~D.}\ \bibnamefont {Tygar}},\ }\bibfield  {title}
  {\enquote {\bibinfo {title} {Adversarial machine learning},}\ }in\ \href
  {https://dl.acm.org/citation.cfm?id=2046692} {\emph {\bibinfo {booktitle}
  {Proceedings of the 4th ACM workshop on Security and artificial
  intelligence}}}\ (\bibinfo {organization} {ACM},\ \bibinfo {year} {2011})\
  pp.\ \bibinfo {pages} {43--58}\BibitemShut {NoStop}%
\bibitem [{\citenamefont {Vorobeychik}\ and\ \citenamefont
  {Kantarcioglu}(2018)}]{vorobeychik2018adversarial}%
  \BibitemOpen
  \bibfield  {author} {\bibinfo {author} {\bibfnamefont {Y.}~\bibnamefont
  {Vorobeychik}}\ and\ \bibinfo {author} {\bibfnamefont {M.}~\bibnamefont
  {Kantarcioglu}},\ }\bibfield  {title} {\enquote {\bibinfo {title}
  {Adversarial machine learning},}\ }\href {\doibase
  10.2200/S00861ED1V01Y201806AIM039} {\bibfield  {journal} {\bibinfo  {journal}
  {Synthesis Lectures on Artificial Intelligence and Machine Learning}\
  }\textbf {\bibinfo {volume} {12}},\ \bibinfo {pages} {1} (\bibinfo {year}
  {2018})}\BibitemShut {NoStop}%
\bibitem [{\citenamefont {Miller}\ \emph {et~al.}(2020)\citenamefont {Miller},
  \citenamefont {Xiang},\ and\ \citenamefont
  {Kesidis}}]{miller2019adversarial}%
  \BibitemOpen
  \bibfield  {author} {\bibinfo {author} {\bibfnamefont {D.~J.}\ \bibnamefont
  {Miller}}, \bibinfo {author} {\bibfnamefont {Z.}~\bibnamefont {Xiang}}, \
  and\ \bibinfo {author} {\bibfnamefont {G.}~\bibnamefont {Kesidis}},\
  }\bibfield  {title} {\enquote {\bibinfo {title} {Adversarial learning
  targeting deep neural network classification: A comprehensive review of
  defenses against attacks},}\ }\href {\doibase 10.1109/JPROC.2020.2970615}
  {\bibfield  {journal} {\bibinfo  {journal} {Proceedings of the IEEE}\
  }\textbf {\bibinfo {volume} {108}},\ \bibinfo {pages} {402} (\bibinfo {year}
  {2020})}\BibitemShut {NoStop}%
\bibitem [{\citenamefont {Goodfellow}\ \emph {et~al.}(2015)\citenamefont
  {Goodfellow}, \citenamefont {Shlens},\ and\ \citenamefont
  {Szegedy}}]{goodfellow2014explaining}%
  \BibitemOpen
  \bibfield  {author} {\bibinfo {author} {\bibfnamefont {I.}~\bibnamefont
  {Goodfellow}}, \bibinfo {author} {\bibfnamefont {J.}~\bibnamefont {Shlens}},
  \ and\ \bibinfo {author} {\bibfnamefont {C.}~\bibnamefont {Szegedy}},\
  }\bibfield  {title} {\enquote {\bibinfo {title} {Explaining and harnessing
  adversarial examples},}\ }in\ \href {http://arxiv.org/abs/1412.6572} {\emph
  {\bibinfo {booktitle} {International Conference on Learning
  Representations}}}\ (\bibinfo {year} {2015})\BibitemShut {NoStop}%
\bibitem [{\citenamefont {Liu}\ and\ \citenamefont
  {Wittek}(2020)}]{liu2019vulnerability}%
  \BibitemOpen
  \bibfield  {author} {\bibinfo {author} {\bibfnamefont {N.}~\bibnamefont
  {Liu}}\ and\ \bibinfo {author} {\bibfnamefont {P.}~\bibnamefont {Wittek}},\
  }\bibfield  {title} {\enquote {\bibinfo {title} {Vulnerability of quantum
  classification to adversarial perturbations},}\ }\href {\doibase
  10.1103/PhysRevA.101.062331} {\bibfield  {journal} {\bibinfo  {journal}
  {Phys. Rev. A}\ }\textbf {\bibinfo {volume} {101}},\ \bibinfo {pages}
  {062331} (\bibinfo {year} {2020})}\BibitemShut {NoStop}%
\bibitem [{\citenamefont {Schmidt}\ \emph {et~al.}(2018)\citenamefont
  {Schmidt}, \citenamefont {Santurkar}, \citenamefont {Tsipras}, \citenamefont
  {Talwar},\ and\ \citenamefont {Madry}}]{Schmidt2018Adv}%
  \BibitemOpen
  \bibfield  {author} {\bibinfo {author} {\bibfnamefont {L.}~\bibnamefont
  {Schmidt}}, \bibinfo {author} {\bibfnamefont {S.}~\bibnamefont {Santurkar}},
  \bibinfo {author} {\bibfnamefont {D.}~\bibnamefont {Tsipras}}, \bibinfo
  {author} {\bibfnamefont {K.}~\bibnamefont {Talwar}}, \ and\ \bibinfo {author}
  {\bibfnamefont {A.}~\bibnamefont {Madry}},\ }\bibfield  {title} {\enquote
  {\bibinfo {title} {Adversarially robust generalization requires more data},}\
  }in\ \href
  {https://proceedings.neurips.cc/paper/2018/file/f708f064faaf32a43e4d3c784e6af9ea-Paper.pdf}
  {\emph {\bibinfo {booktitle} {Advances in Neural Information Processing
  Systems}}}\ (\bibinfo {year} {2018})\BibitemShut {NoStop}%
\bibitem [{\citenamefont {Szegedy}\ \emph {et~al.}(2014)\citenamefont
  {Szegedy}, \citenamefont {Zaremba}, \citenamefont {Sutskever}, \citenamefont
  {Bruna}, \citenamefont {Erhan}, \citenamefont {Goodfellow},\ and\
  \citenamefont {Fergus}}]{Szegedy2013Intriguing}%
  \BibitemOpen
  \bibfield  {author} {\bibinfo {author} {\bibfnamefont {C.}~\bibnamefont
  {Szegedy}}, \bibinfo {author} {\bibfnamefont {W.}~\bibnamefont {Zaremba}},
  \bibinfo {author} {\bibfnamefont {I.}~\bibnamefont {Sutskever}}, \bibinfo
  {author} {\bibfnamefont {J.}~\bibnamefont {Bruna}}, \bibinfo {author}
  {\bibfnamefont {D.}~\bibnamefont {Erhan}}, \bibinfo {author} {\bibfnamefont
  {I.~J.}\ \bibnamefont {Goodfellow}}, \ and\ \bibinfo {author} {\bibfnamefont
  {R.}~\bibnamefont {Fergus}},\ }\bibfield  {title} {\enquote {\bibinfo {title}
  {Intriguing properties of neural networks},}\ }in\ \href
  {http://arxiv.org/abs/1312.6199} {\emph {\bibinfo {booktitle} {International
  Conference on Learning Representations}}}\ (\bibinfo {year}
  {2014})\BibitemShut {NoStop}%
\bibitem [{\citenamefont {Zhou}\ \emph {et~al.}(2015)\citenamefont {Zhou},
  \citenamefont {Khosla}, \citenamefont {Lapedriza}, \citenamefont {Oliva},\
  and\ \citenamefont {Torralba}}]{zhou2015object}%
  \BibitemOpen
  \bibfield  {author} {\bibinfo {author} {\bibfnamefont {B.}~\bibnamefont
  {Zhou}}, \bibinfo {author} {\bibfnamefont {A.}~\bibnamefont {Khosla}},
  \bibinfo {author} {\bibfnamefont {{\`{A}}.}~\bibnamefont {Lapedriza}},
  \bibinfo {author} {\bibfnamefont {A.}~\bibnamefont {Oliva}}, \ and\ \bibinfo
  {author} {\bibfnamefont {A.}~\bibnamefont {Torralba}},\ }\bibfield  {title}
  {\enquote {\bibinfo {title} {Object detectors emerge in deep scene cnns},}\
  }in\ \href {http://arxiv.org/abs/1412.6856} {\emph {\bibinfo {booktitle}
  {International Conference on Learning Representations}}}\ (\bibinfo {year}
  {2015})\BibitemShut {NoStop}%
\bibitem [{\citenamefont {Zhou}\ \emph {et~al.}(2016)\citenamefont {Zhou},
  \citenamefont {Khosla}, \citenamefont {Lapedriza}, \citenamefont {Oliva},\
  and\ \citenamefont {Torralba}}]{zhou2015learning}%
  \BibitemOpen
  \bibfield  {author} {\bibinfo {author} {\bibfnamefont {B.}~\bibnamefont
  {Zhou}}, \bibinfo {author} {\bibfnamefont {A.}~\bibnamefont {Khosla}},
  \bibinfo {author} {\bibfnamefont {{\`{A}}.}~\bibnamefont {Lapedriza}},
  \bibinfo {author} {\bibfnamefont {A.}~\bibnamefont {Oliva}}, \ and\ \bibinfo
  {author} {\bibfnamefont {A.}~\bibnamefont {Torralba}},\ }\bibfield  {title}
  {\enquote {\bibinfo {title} {Learning deep features for discriminative
  localization},}\ }in\ \href {https://doi.org/10.1109/CVPR.2016.319} {\emph
  {\bibinfo {booktitle} {{IEEE} Conference on Computer Vision and Pattern
  Recognition}}}\ (\bibinfo {year} {2016})\ pp.\ \bibinfo {pages}
  {2921--2929}\BibitemShut {NoStop}%
\bibitem [{VML()}]{VMLPMSup}%
  \BibitemOpen
  \href@noop {} {}\bibinfo {note} {See Supplemental Material at [URL will be
  inserted by publisher] for details on different methods to obtain adversarial
  examples, the derivation and explaination of activation maps, and for more
  numerical results on both adversarial examples generation and adversarial
  training.}\BibitemShut {Stop}%
\bibitem [{\citenamefont {Storn}\ and\ \citenamefont
  {Price}(1997)}]{storn1997differential}%
  \BibitemOpen
  \bibfield  {author} {\bibinfo {author} {\bibfnamefont {R.}~\bibnamefont
  {Storn}}\ and\ \bibinfo {author} {\bibfnamefont {K.}~\bibnamefont {Price}},\
  }\bibfield  {title} {\enquote {\bibinfo {title} {Differential evolution--a
  simple and efficient heuristic for global optimization over continuous
  spaces},}\ }\href {https://link.springer.com/article/10.1023/A:1008202821328}
  {\bibfield  {journal} {\bibinfo  {journal} {Journal of global optimization}\
  }\textbf {\bibinfo {volume} {11}},\ \bibinfo {pages} {341} (\bibinfo {year}
  {1997})}\BibitemShut {NoStop}%
\bibitem [{\citenamefont {Das}\ and\ \citenamefont
  {Suganthan}(2010)}]{das2010differential}%
  \BibitemOpen
  \bibfield  {author} {\bibinfo {author} {\bibfnamefont {S.}~\bibnamefont
  {Das}}\ and\ \bibinfo {author} {\bibfnamefont {P.~N.}\ \bibnamefont
  {Suganthan}},\ }\bibfield  {title} {\enquote {\bibinfo {title} {Differential
  evolution: A survey of the state-of-the-art},}\ }\href {\doibase
  10.1109/TEVC.2010.2059031} {\bibfield  {journal} {\bibinfo  {journal} {IEEE
  Transactions on Evolutionary Computation}\ }\textbf {\bibinfo {volume}
  {15}},\ \bibinfo {pages} {4} (\bibinfo {year} {2010})}\BibitemShut {NoStop}%
\bibitem [{\citenamefont {Madry}\ \emph {et~al.}(2019)\citenamefont {Madry},
  \citenamefont {Makelov}, \citenamefont {Schmidt}, \citenamefont {Tsipras},\
  and\ \citenamefont {Vladu}}]{Madry2017Towards}%
  \BibitemOpen
  \bibfield  {author} {\bibinfo {author} {\bibfnamefont {A.}~\bibnamefont
  {Madry}}, \bibinfo {author} {\bibfnamefont {A.}~\bibnamefont {Makelov}},
  \bibinfo {author} {\bibfnamefont {L.}~\bibnamefont {Schmidt}}, \bibinfo
  {author} {\bibfnamefont {D.}~\bibnamefont {Tsipras}}, \ and\ \bibinfo
  {author} {\bibfnamefont {A.}~\bibnamefont {Vladu}},\ }\href@noop {} {\enquote
  {\bibinfo {title} {Towards deep learning models resistant to adversarial
  attacks},}\ } (\bibinfo {year} {2019}),\ \Eprint
  {http://arxiv.org/abs/1706.06083} {arXiv:1706.06083 [stat.ML]} \BibitemShut
  {NoStop}%
\bibitem [{\citenamefont {Dong}\ \emph {et~al.}(2018)\citenamefont {Dong},
  \citenamefont {Liao}, \citenamefont {Pang}, \citenamefont {Su}, \citenamefont
  {Zhu}, \citenamefont {Hu},\ and\ \citenamefont {Li}}]{dong2018boosting}%
  \BibitemOpen
  \bibfield  {author} {\bibinfo {author} {\bibfnamefont {Y.}~\bibnamefont
  {Dong}}, \bibinfo {author} {\bibfnamefont {F.}~\bibnamefont {Liao}}, \bibinfo
  {author} {\bibfnamefont {T.}~\bibnamefont {Pang}}, \bibinfo {author}
  {\bibfnamefont {H.}~\bibnamefont {Su}}, \bibinfo {author} {\bibfnamefont
  {J.}~\bibnamefont {Zhu}}, \bibinfo {author} {\bibfnamefont {X.}~\bibnamefont
  {Hu}}, \ and\ \bibinfo {author} {\bibfnamefont {J.}~\bibnamefont {Li}},\
  }\bibfield  {title} {\enquote {\bibinfo {title} {Boosting adversarial attacks
  with momentum},}\ }in\ \href
  {http://openaccess.thecvf.com/content_cvpr_2018/html/Dong_Boosting_Adversarial_Attacks_CVPR_2018_paper.html}
  {\emph {\bibinfo {booktitle} {Proceedings of the IEEE conference on computer
  vision and pattern recognition}}}\ (\bibinfo {year} {2018})\ pp.\ \bibinfo
  {pages} {9185--9193}\BibitemShut {NoStop}%
\bibitem [{\citenamefont {Onsager}(1944)}]{Onsager1944Crystal}%
  \BibitemOpen
  \bibfield  {author} {\bibinfo {author} {\bibfnamefont {L.}~\bibnamefont
  {Onsager}},\ }\bibfield  {title} {\enquote {\bibinfo {title} {Crystal
  statistics. i. a two-dimensional model with an order-disorder transition},}\
  }\href {\doibase 10.1103/PhysRev.65.117} {\bibfield  {journal} {\bibinfo
  {journal} {Phys. Rev.}\ }\textbf {\bibinfo {volume} {65}},\ \bibinfo {pages}
  {117} (\bibinfo {year} {1944})}\BibitemShut {NoStop}%
\bibitem [{\citenamefont {Su}\ \emph {et~al.}(2019)\citenamefont {Su},
  \citenamefont {Vargas},\ and\ \citenamefont {Sakurai}}]{su2019one}%
  \BibitemOpen
  \bibfield  {author} {\bibinfo {author} {\bibfnamefont {J.}~\bibnamefont
  {Su}}, \bibinfo {author} {\bibfnamefont {D.~V.}\ \bibnamefont {Vargas}}, \
  and\ \bibinfo {author} {\bibfnamefont {K.}~\bibnamefont {Sakurai}},\
  }\bibfield  {title} {\enquote {\bibinfo {title} {One pixel attack for fooling
  deep neural networks},}\ }\href
  {https://ieeexplore.ieee.org/document/8601309} {\bibfield  {journal}
  {\bibinfo  {journal} {IEEE Transactions on Evolutionary Computation}\ }
  (\bibinfo {year} {2019})}\BibitemShut {NoStop}%
\bibitem [{\citenamefont {Lifshitz}\ and\ \citenamefont
  {Pitaevskii}(2013)}]{lifshitz2013statistical}%
  \BibitemOpen
  \bibfield  {author} {\bibinfo {author} {\bibfnamefont {E.~M.}\ \bibnamefont
  {Lifshitz}}\ and\ \bibinfo {author} {\bibfnamefont {L.~P.}\ \bibnamefont
  {Pitaevskii}},\ }\href@noop {} {\emph {\bibinfo {title} {Statistical physics:
  theory of the condensed state}}},\ Vol.~\bibinfo {volume} {9}\ (\bibinfo
  {publisher} {Elsevier},\ \bibinfo {year} {2013})\BibitemShut {NoStop}%
\bibitem [{\citenamefont {Qi}\ and\ \citenamefont
  {Zhang}(2011)}]{qi2011topological}%
  \BibitemOpen
  \bibfield  {author} {\bibinfo {author} {\bibfnamefont {X.-L.}\ \bibnamefont
  {Qi}}\ and\ \bibinfo {author} {\bibfnamefont {S.-C.}\ \bibnamefont {Zhang}},\
  }\bibfield  {title} {\enquote {\bibinfo {title} {Topological insulators and
  superconductors},}\ }\href {\doibase 10.1103/RevModPhys.83.1057} {\bibfield
  {journal} {\bibinfo  {journal} {Rev. Mod. Phys.}\ }\textbf {\bibinfo {volume}
  {83}},\ \bibinfo {pages} {1057} (\bibinfo {year} {2011})}\BibitemShut
  {NoStop}%
\bibitem [{\citenamefont {Hasan}\ and\ \citenamefont
  {Kane}(2010)}]{hasan2010colloquium}%
  \BibitemOpen
  \bibfield  {author} {\bibinfo {author} {\bibfnamefont {M.~Z.}\ \bibnamefont
  {Hasan}}\ and\ \bibinfo {author} {\bibfnamefont {C.~L.}\ \bibnamefont
  {Kane}},\ }\bibfield  {title} {\enquote {\bibinfo {title} {Colloquium:
  Topological insulators},}\ }\href {\doibase 10.1103/RevModPhys.82.3045}
  {\bibfield  {journal} {\bibinfo  {journal} {Rev. Mod. Phys.}\ }\textbf
  {\bibinfo {volume} {82}},\ \bibinfo {pages} {3045} (\bibinfo {year}
  {2010})}\BibitemShut {NoStop}%
\bibitem [{\citenamefont {Zhang}\ \emph {et~al.}(2017)\citenamefont {Zhang},
  \citenamefont {Melko},\ and\ \citenamefont {Kim}}]{Zhang2017Machine}%
  \BibitemOpen
  \bibfield  {author} {\bibinfo {author} {\bibfnamefont {Y.}~\bibnamefont
  {Zhang}}, \bibinfo {author} {\bibfnamefont {R.~G.}\ \bibnamefont {Melko}}, \
  and\ \bibinfo {author} {\bibfnamefont {E.-A.}\ \bibnamefont {Kim}},\
  }\bibfield  {title} {\enquote {\bibinfo {title} {Machine learning $z_2$
  quantum spin liquids with quasiparticle statistics},}\ }\href {\doibase
  10.1103/PhysRevB.96.245119} {\bibfield  {journal} {\bibinfo  {journal} {Phys.
  Rev. B}\ }\textbf {\bibinfo {volume} {96}},\ \bibinfo {pages} {245119}
  (\bibinfo {year} {2017})}\BibitemShut {NoStop}%
\bibitem [{\citenamefont {Yu}\ and\ \citenamefont
  {Deng}(2021)}]{Yu2021Unsupervised}%
  \BibitemOpen
  \bibfield  {author} {\bibinfo {author} {\bibfnamefont {L.-W.}\ \bibnamefont
  {Yu}}\ and\ \bibinfo {author} {\bibfnamefont {D.-L.}\ \bibnamefont {Deng}},\
  }\bibfield  {title} {\enquote {\bibinfo {title} {Unsupervised learning of
  non-hermitian topological phases},}\ }\href
  {http://dx.doi.org/10.1103/PhysRevLett.126.240402} {\bibfield  {journal}
  {\bibinfo  {journal} {Phys. Rev. Lett.}\ }\textbf {\bibinfo {volume} {126}}
  (\bibinfo {year} {2021})}\BibitemShut {NoStop}%
\bibitem [{\citenamefont {Scheurer}\ and\ \citenamefont
  {Slager}(2020)}]{Scheurer2020Unsupervised}%
  \BibitemOpen
  \bibfield  {author} {\bibinfo {author} {\bibfnamefont {M.~S.}\ \bibnamefont
  {Scheurer}}\ and\ \bibinfo {author} {\bibfnamefont {R.-J.}\ \bibnamefont
  {Slager}},\ }\bibfield  {title} {\enquote {\bibinfo {title} {Unsupervised
  machine learning and band topology},}\ }\href
  {http://dx.doi.org/10.1103/PhysRevLett.124.226401} {\bibfield  {journal}
  {\bibinfo  {journal} {Phys. Rev. Lett.}\ }\textbf {\bibinfo {volume} {124}}
  (\bibinfo {year} {2020})}\BibitemShut {NoStop}%
\bibitem [{\citenamefont {Long}\ \emph {et~al.}(2020)\citenamefont {Long},
  \citenamefont {Ren},\ and\ \citenamefont {Chen}}]{Long2020Unsupervised}%
  \BibitemOpen
  \bibfield  {author} {\bibinfo {author} {\bibfnamefont {Y.}~\bibnamefont
  {Long}}, \bibinfo {author} {\bibfnamefont {J.}~\bibnamefont {Ren}}, \ and\
  \bibinfo {author} {\bibfnamefont {H.}~\bibnamefont {Chen}},\ }\bibfield
  {title} {\enquote {\bibinfo {title} {Unsupervised manifold clustering of
  topological phononics},}\ }\href
  {http://dx.doi.org/10.1103/PhysRevLett.124.185501} {\bibfield  {journal}
  {\bibinfo  {journal} {Phys. Rev. Lett.}\ }\textbf {\bibinfo {volume} {124}}
  (\bibinfo {year} {2020})}\BibitemShut {NoStop}%
\bibitem [{\citenamefont {Lidiak}\ and\ \citenamefont
  {Gong}(2020)}]{Lidiak2020Unsupervised}%
  \BibitemOpen
  \bibfield  {author} {\bibinfo {author} {\bibfnamefont {A.}~\bibnamefont
  {Lidiak}}\ and\ \bibinfo {author} {\bibfnamefont {Z.}~\bibnamefont {Gong}},\
  }\bibfield  {title} {\enquote {\bibinfo {title} {Unsupervised machine
  learning of quantum phase transitions using diffusion maps},}\ }\href
  {http://dx.doi.org/10.1103/PhysRevLett.125.225701} {\bibfield  {journal}
  {\bibinfo  {journal} {Phys. Rev. Lett.}\ }\textbf {\bibinfo {volume} {125}}
  (\bibinfo {year} {2020})}\BibitemShut {NoStop}%
\bibitem [{\citenamefont {Che}\ \emph {et~al.}(2020)\citenamefont {Che},
  \citenamefont {Gneiting}, \citenamefont {Liu},\ and\ \citenamefont
  {Nori}}]{Che2020Topological}%
  \BibitemOpen
  \bibfield  {author} {\bibinfo {author} {\bibfnamefont {Y.}~\bibnamefont
  {Che}}, \bibinfo {author} {\bibfnamefont {C.}~\bibnamefont {Gneiting}},
  \bibinfo {author} {\bibfnamefont {T.}~\bibnamefont {Liu}}, \ and\ \bibinfo
  {author} {\bibfnamefont {F.}~\bibnamefont {Nori}},\ }\bibfield  {title}
  {\enquote {\bibinfo {title} {Topological quantum phase transitions retrieved
  through unsupervised machine learning},}\ }\href
  {http://dx.doi.org/10.1103/PhysRevB.102.134213} {\bibfield  {journal}
  {\bibinfo  {journal} {Phys. Rev. B}\ }\textbf {\bibinfo {volume} {102}}
  (\bibinfo {year} {2020})}\BibitemShut {NoStop}%
\bibitem [{\citenamefont {Fukushima}\ \emph {et~al.}(2019)\citenamefont
  {Fukushima}, \citenamefont {Funai},\ and\ \citenamefont
  {Iida}}]{fukushima2019featuring}%
  \BibitemOpen
  \bibfield  {author} {\bibinfo {author} {\bibfnamefont {K.}~\bibnamefont
  {Fukushima}}, \bibinfo {author} {\bibfnamefont {S.~S.}\ \bibnamefont
  {Funai}}, \ and\ \bibinfo {author} {\bibfnamefont {H.}~\bibnamefont {Iida}},\
  }\href@noop {} {\enquote {\bibinfo {title} {Featuring the topology with the
  unsupervised machine learning},}\ } (\bibinfo {year} {2019}),\ \Eprint
  {http://arxiv.org/abs/1908.00281} {arXiv:1908.00281 [cs.LG]} \BibitemShut
  {NoStop}%
\bibitem [{\citenamefont {Sch\"afer}\ and\ \citenamefont
  {L\"orch}(2019)}]{Schafer2019Vector}%
  \BibitemOpen
  \bibfield  {author} {\bibinfo {author} {\bibfnamefont {F.}~\bibnamefont
  {Sch\"afer}}\ and\ \bibinfo {author} {\bibfnamefont {N.}~\bibnamefont
  {L\"orch}},\ }\bibfield  {title} {\enquote {\bibinfo {title} {Vector field
  divergence of predictive model output as indication of phase transitions},}\
  }\href {https://link.aps.org/doi/10.1103/PhysRevE.99.062107} {\bibfield
  {journal} {\bibinfo  {journal} {Phys. Rev. E}\ }\textbf {\bibinfo {volume}
  {99}},\ \bibinfo {pages} {062107} (\bibinfo {year} {2019})}\BibitemShut
  {NoStop}%
\bibitem [{\citenamefont {Balabanov}\ and\ \citenamefont
  {Granath}(2020)}]{Balabanov2020Unsupervised}%
  \BibitemOpen
  \bibfield  {author} {\bibinfo {author} {\bibfnamefont {O.}~\bibnamefont
  {Balabanov}}\ and\ \bibinfo {author} {\bibfnamefont {M.}~\bibnamefont
  {Granath}},\ }\bibfield  {title} {\enquote {\bibinfo {title} {Unsupervised
  learning using topological data augmentation},}\ }\href
  {http://dx.doi.org/10.1103/PhysRevResearch.2.013354} {\bibfield  {journal}
  {\bibinfo  {journal} {Phys. Rev. Research}\ }\textbf {\bibinfo {volume} {2}}
  (\bibinfo {year} {2020})}\BibitemShut {NoStop}%
\bibitem [{\citenamefont {Alexandrou}\ \emph {et~al.}(2020)\citenamefont
  {Alexandrou}, \citenamefont {Athenodorou}, \citenamefont {Chrysostomou},\
  and\ \citenamefont {Paul}}]{Alexandrou2020The}%
  \BibitemOpen
  \bibfield  {author} {\bibinfo {author} {\bibfnamefont {C.}~\bibnamefont
  {Alexandrou}}, \bibinfo {author} {\bibfnamefont {A.}~\bibnamefont
  {Athenodorou}}, \bibinfo {author} {\bibfnamefont {C.}~\bibnamefont
  {Chrysostomou}}, \ and\ \bibinfo {author} {\bibfnamefont {S.}~\bibnamefont
  {Paul}},\ }\bibfield  {title} {\enquote {\bibinfo {title} {The critical
  temperature of the 2d-ising model through deep learning autoencoders},}\
  }\href {\doibase 10.1140/epjb/e2020-100506-5} {\bibfield  {journal} {\bibinfo
   {journal} {Eur. Phys. J. B}\ }\textbf {\bibinfo {volume} {93}},\ \bibinfo
  {pages} {226} (\bibinfo {year} {2020})}\BibitemShut {NoStop}%
\bibitem [{\citenamefont {Greplov{\'a}}\ \emph {et~al.}(2020)\citenamefont
  {Greplov{\'a}}, \citenamefont {Valenti}, \citenamefont {Boschung},
  \citenamefont {Schafer}, \citenamefont {Lorch},\ and\ \citenamefont
  {Huber}}]{Greplov2019nsupervised}%
  \BibitemOpen
  \bibfield  {author} {\bibinfo {author} {\bibfnamefont {E.}~\bibnamefont
  {Greplov{\'a}}}, \bibinfo {author} {\bibfnamefont {A.}~\bibnamefont
  {Valenti}}, \bibinfo {author} {\bibfnamefont {G.}~\bibnamefont {Boschung}},
  \bibinfo {author} {\bibfnamefont {F.}~\bibnamefont {Schafer}}, \bibinfo
  {author} {\bibfnamefont {N.}~\bibnamefont {Lorch}}, \ and\ \bibinfo {author}
  {\bibfnamefont {S.}~\bibnamefont {Huber}},\ }\bibfield  {title} {\enquote
  {\bibinfo {title} {Unsupervised identification of topological order using
  predictive models},}\ }\href {\doibase 10.1088/1367-2630/ab7771} {\bibfield
  {journal} {\bibinfo  {journal} {New Journal of Physics}\ }\textbf {\bibinfo
  {volume} {22}},\ \bibinfo {pages} {045003} (\bibinfo {year}
  {2020})}\BibitemShut {NoStop}%
\bibitem [{\citenamefont {Arnold}\ \emph {et~al.}(2021)\citenamefont {Arnold},
  \citenamefont {Schäfer}, \citenamefont {Žonda},\ and\ \citenamefont
  {Lode}}]{Arnold2021Interpretable}%
  \BibitemOpen
  \bibfield  {author} {\bibinfo {author} {\bibfnamefont {J.}~\bibnamefont
  {Arnold}}, \bibinfo {author} {\bibfnamefont {F.}~\bibnamefont {Schäfer}},
  \bibinfo {author} {\bibfnamefont {M.}~\bibnamefont {Žonda}}, \ and\ \bibinfo
  {author} {\bibfnamefont {A.~U.~J.}\ \bibnamefont {Lode}},\ }\bibfield
  {title} {\enquote {\bibinfo {title} {Interpretable and unsupervised phase
  classification},}\ }\href
  {http://dx.doi.org/10.1103/PhysRevResearch.3.033052} {\bibfield  {journal}
  {\bibinfo  {journal} {Phys. Rev. Research}\ }\textbf {\bibinfo {volume} {3}}
  (\bibinfo {year} {2021})}\BibitemShut {NoStop}%
\bibitem [{\citenamefont {Neupert}\ \emph {et~al.}(2012)\citenamefont
  {Neupert}, \citenamefont {Santos}, \citenamefont {Ryu}, \citenamefont
  {Chamon},\ and\ \citenamefont {Mudry}}]{neupert2012noncommutative}%
  \BibitemOpen
  \bibfield  {author} {\bibinfo {author} {\bibfnamefont {T.}~\bibnamefont
  {Neupert}}, \bibinfo {author} {\bibfnamefont {L.}~\bibnamefont {Santos}},
  \bibinfo {author} {\bibfnamefont {S.}~\bibnamefont {Ryu}}, \bibinfo {author}
  {\bibfnamefont {C.}~\bibnamefont {Chamon}}, \ and\ \bibinfo {author}
  {\bibfnamefont {C.}~\bibnamefont {Mudry}},\ }\bibfield  {title} {\enquote
  {\bibinfo {title} {Noncommutative geometry for three-dimensional topological
  insulators},}\ }\href {\doibase 10.1103/PhysRevB.86.035125} {\bibfield
  {journal} {\bibinfo  {journal} {Phys. Rev. B}\ }\textbf {\bibinfo {volume}
  {86}},\ \bibinfo {pages} {035125} (\bibinfo {year} {2012})}\BibitemShut
  {NoStop}%
\bibitem [{\citenamefont {Wang}\ \emph {et~al.}(2014)\citenamefont {Wang},
  \citenamefont {Deng},\ and\ \citenamefont {Duan}}]{wang2014probe}%
  \BibitemOpen
  \bibfield  {author} {\bibinfo {author} {\bibfnamefont {S.-T.}\ \bibnamefont
  {Wang}}, \bibinfo {author} {\bibfnamefont {D.-L.}\ \bibnamefont {Deng}}, \
  and\ \bibinfo {author} {\bibfnamefont {L.-M.}\ \bibnamefont {Duan}},\
  }\bibfield  {title} {\enquote {\bibinfo {title} {Probe of three-dimensional
  chiral topological insulators in an optical lattice},}\ }\href {\doibase
  10.1103/PhysRevLett.113.033002} {\bibfield  {journal} {\bibinfo  {journal}
  {Phys. Rev. Lett.}\ }\textbf {\bibinfo {volume} {113}},\ \bibinfo {pages}
  {033002} (\bibinfo {year} {2014})}\BibitemShut {NoStop}%
\bibitem [{Top()}]{TopInv}%
  \BibitemOpen
  \href@noop {} {}\bibinfo {note} {For each band, the topological invariant
  $\chi^{(\eta)}$ (here $\eta = l,m,u$ denotes the lower, middle and upper
  bands, respectively) can be written as an integral in the 3D momentum space:
  \begin{eqnarray*} \chi^{(\eta)} & = &
  \frac{1}{4\pi^2}\int_{\text{BZ}}\epsilon^{\mu\nu\tau}A_{\mu}^{(\eta)}\partial_{k^{\nu}}A_{\tau}^{(\eta)}d^{3}\mathbf{k},
  \end{eqnarray*} where $\epsilon^{\mu\nu\tau}$ is the Levi-Civita symbol with
  $\mu,\nu,\tau\in\{x,y,z\}$, and the Berry connection is
  $A_{\mu}^{(\eta)}=\langle\psi_{\mathbf{k}}^{(\eta)}|\partial_{k^{\mu}}|\psi_{\mathbf{k}}^{(\eta)}\rangle$
  with $|\psi_{\mathbf{k}}^{(\eta)}\rangle$ denoting the Bloch state for the
  $\eta$ band. The topological invariants for each band are related as
  $\chi^{(u)} = \chi^{(l)} = \chi^{(m)}/4$ for the three-band chiral
  topological insulator model studied in this paper.}\BibitemShut {Stop}%
\bibitem [{\citenamefont {Papernot}\ \emph {et~al.}(2018)\citenamefont
  {Papernot}, \citenamefont {Faghri}, \citenamefont {Carlini}, \citenamefont
  {Goodfellow}, \citenamefont {Feinman}, \citenamefont {Kurakin}, \citenamefont
  {Xie}, \citenamefont {Sharma}, \citenamefont {Brown}, \citenamefont {Roy}
  \emph {et~al.}}]{papernot2018cleverhans}%
  \BibitemOpen
  \bibfield  {author} {\bibinfo {author} {\bibfnamefont {N.}~\bibnamefont
  {Papernot}}, \bibinfo {author} {\bibfnamefont {F.}~\bibnamefont {Faghri}},
  \bibinfo {author} {\bibfnamefont {N.}~\bibnamefont {Carlini}}, \bibinfo
  {author} {\bibfnamefont {I.}~\bibnamefont {Goodfellow}}, \bibinfo {author}
  {\bibfnamefont {R.}~\bibnamefont {Feinman}}, \bibinfo {author} {\bibfnamefont
  {A.}~\bibnamefont {Kurakin}}, \bibinfo {author} {\bibfnamefont
  {C.}~\bibnamefont {Xie}}, \bibinfo {author} {\bibfnamefont {Y.}~\bibnamefont
  {Sharma}}, \bibinfo {author} {\bibfnamefont {T.}~\bibnamefont {Brown}},
  \bibinfo {author} {\bibfnamefont {A.}~\bibnamefont {Roy}},  \emph {et~al.},\
  }\href@noop {} {\enquote {\bibinfo {title} {Technical report on the
  cleverhans v2.1.0 adversarial examples library},}\ } (\bibinfo {year}
  {2018}),\ \Eprint {http://arxiv.org/abs/1610.00768} {arXiv:1610.00768
  [cs.LG]} \BibitemShut {NoStop}%
\bibitem [{\citenamefont {Tsipras}\ \emph {et~al.}(2019)\citenamefont
  {Tsipras}, \citenamefont {Santurkar}, \citenamefont {Engstrom}, \citenamefont
  {Turner},\ and\ \citenamefont {Madry}}]{tsipras2018robustness}%
  \BibitemOpen
  \bibfield  {author} {\bibinfo {author} {\bibfnamefont {D.}~\bibnamefont
  {Tsipras}}, \bibinfo {author} {\bibfnamefont {S.}~\bibnamefont {Santurkar}},
  \bibinfo {author} {\bibfnamefont {L.}~\bibnamefont {Engstrom}}, \bibinfo
  {author} {\bibfnamefont {A.}~\bibnamefont {Turner}}, \ and\ \bibinfo {author}
  {\bibfnamefont {A.}~\bibnamefont {Madry}},\ }\bibfield  {title} {\enquote
  {\bibinfo {title} {Robustness may be at odds with accuracy},}\ }in\ \href
  {https://openreview.net/forum?id=SyxAb30cY7} {\emph {\bibinfo {booktitle}
  {International Conference on Learning Representations}}}\ (\bibinfo {year}
  {2019})\BibitemShut {NoStop}%
\bibitem [{\citenamefont {Fawzi}\ \emph {et~al.}(2018)\citenamefont {Fawzi},
  \citenamefont {Fawzi},\ and\ \citenamefont {Fawzi}}]{fawzi2018adversarial}%
  \BibitemOpen
  \bibfield  {author} {\bibinfo {author} {\bibfnamefont {A.}~\bibnamefont
  {Fawzi}}, \bibinfo {author} {\bibfnamefont {H.}~\bibnamefont {Fawzi}}, \ and\
  \bibinfo {author} {\bibfnamefont {O.}~\bibnamefont {Fawzi}},\ }\bibfield
  {title} {\enquote {\bibinfo {title} {Adversarial vulnerability for any
  classifier},}\ }in\ \href
  {http://papers.nips.cc/paper/7394-adversarial-vulnerability-for-any-classifier}
  {\emph {\bibinfo {booktitle} {Advances in Neural Information Processing
  Systems}}}\ (\bibinfo {year} {2018})\ pp.\ \bibinfo {pages}
  {1178--1187}\BibitemShut {NoStop}%
\bibitem [{\citenamefont {Gilmer}\ \emph {et~al.}(2018)\citenamefont {Gilmer},
  \citenamefont {Metz}, \citenamefont {Faghri}, \citenamefont {Schoenholz},
  \citenamefont {Raghu}, \citenamefont {Wattenberg},\ and\ \citenamefont
  {Goodfellow}}]{gilmer2018adversarial}%
  \BibitemOpen
  \bibfield  {author} {\bibinfo {author} {\bibfnamefont {J.}~\bibnamefont
  {Gilmer}}, \bibinfo {author} {\bibfnamefont {L.}~\bibnamefont {Metz}},
  \bibinfo {author} {\bibfnamefont {F.}~\bibnamefont {Faghri}}, \bibinfo
  {author} {\bibfnamefont {S.~S.}\ \bibnamefont {Schoenholz}}, \bibinfo
  {author} {\bibfnamefont {M.}~\bibnamefont {Raghu}}, \bibinfo {author}
  {\bibfnamefont {M.}~\bibnamefont {Wattenberg}}, \ and\ \bibinfo {author}
  {\bibfnamefont {I.}~\bibnamefont {Goodfellow}},\ }\href@noop {} {\enquote
  {\bibinfo {title} {Adversarial spheres},}\ } (\bibinfo {year} {2018}),\
  \Eprint {http://arxiv.org/abs/1801.02774} {arXiv:1801.02774 [cs.CV]}
  \BibitemShut {NoStop}%
\bibitem [{\citenamefont {Dohmatob}(2019)}]{dohmatob2018limitations}%
  \BibitemOpen
  \bibfield  {author} {\bibinfo {author} {\bibfnamefont {E.}~\bibnamefont
  {Dohmatob}},\ }\bibfield  {title} {\enquote {\bibinfo {title} {Generalized no
  free lunch theorem for adversarial robustness},}\ }in\ \href
  {https://proceedings.mlr.press/v97/dohmatob19a.html} {\emph {\bibinfo
  {booktitle} {Proceedings of the 36th International Conference on Machine
  Learning}}}\ (\bibinfo {year} {2019})\ pp.\ \bibinfo {pages}
  {1646--1654}\BibitemShut {NoStop}%
\bibitem [{\citenamefont {Chakraborty}\ \emph {et~al.}(2018)\citenamefont
  {Chakraborty}, \citenamefont {Alam}, \citenamefont {Dey}, \citenamefont
  {Chattopadhyay},\ and\ \citenamefont
  {Mukhopadhyay}}]{chakraborty2018adversarial}%
  \BibitemOpen
  \bibfield  {author} {\bibinfo {author} {\bibfnamefont {A.}~\bibnamefont
  {Chakraborty}}, \bibinfo {author} {\bibfnamefont {M.}~\bibnamefont {Alam}},
  \bibinfo {author} {\bibfnamefont {V.}~\bibnamefont {Dey}}, \bibinfo {author}
  {\bibfnamefont {A.}~\bibnamefont {Chattopadhyay}}, \ and\ \bibinfo {author}
  {\bibfnamefont {D.}~\bibnamefont {Mukhopadhyay}},\ }\href@noop {} {\enquote
  {\bibinfo {title} {Adversarial attacks and defences: A survey},}\ } (\bibinfo
  {year} {2018}),\ \Eprint {http://arxiv.org/abs/1810.00069} {arXiv:1810.00069
  [cs.LG]} \BibitemShut {NoStop}%
\bibitem [{\citenamefont {Yuan}\ \emph {et~al.}(2019)\citenamefont {Yuan},
  \citenamefont {He}, \citenamefont {Zhu},\ and\ \citenamefont
  {Li}}]{yuan2019adversarial}%
  \BibitemOpen
  \bibfield  {author} {\bibinfo {author} {\bibfnamefont {X.}~\bibnamefont
  {Yuan}}, \bibinfo {author} {\bibfnamefont {P.}~\bibnamefont {He}}, \bibinfo
  {author} {\bibfnamefont {Q.}~\bibnamefont {Zhu}}, \ and\ \bibinfo {author}
  {\bibfnamefont {X.}~\bibnamefont {Li}},\ }\bibfield  {title} {\enquote
  {\bibinfo {title} {Adversarial examples: Attacks and defenses for deep
  learning},}\ }\href {https://ieeexplore.ieee.org/abstract/document/8611298}
  {\bibfield  {journal} {\bibinfo  {journal} {IEEE Transactions on Neural
  Networks and Learning Systems}\ } (\bibinfo {year} {2019})}\BibitemShut
  {NoStop}%
\bibitem [{\citenamefont {Narodytska}\ and\ \citenamefont
  {Kasiviswanathan}(2017)}]{narodytska2017simple}%
  \BibitemOpen
  \bibfield  {author} {\bibinfo {author} {\bibfnamefont {N.}~\bibnamefont
  {Narodytska}}\ and\ \bibinfo {author} {\bibfnamefont {S.}~\bibnamefont
  {Kasiviswanathan}},\ }\bibfield  {title} {\enquote {\bibinfo {title} {Simple
  black-box adversarial attacks on deep neural networks},}\ }in\ \href
  {10.1109/CVPRW.2017.172} {\emph {\bibinfo {booktitle} {2017 IEEE Conference
  on Computer Vision and Pattern Recognition Workshops (CVPRW)}}}\ (\bibinfo
  {organization} {IEEE},\ \bibinfo {year} {2017})\ pp.\ \bibinfo {pages}
  {1310--1318}\BibitemShut {NoStop}%
\bibitem [{\citenamefont {Papernot}\ \emph {et~al.}(2016)\citenamefont
  {Papernot}, \citenamefont {McDaniel},\ and\ \citenamefont
  {Goodfellow}}]{papernot2016transferability}%
  \BibitemOpen
  \bibfield  {author} {\bibinfo {author} {\bibfnamefont {N.}~\bibnamefont
  {Papernot}}, \bibinfo {author} {\bibfnamefont {P.}~\bibnamefont {McDaniel}},
  \ and\ \bibinfo {author} {\bibfnamefont {I.}~\bibnamefont {Goodfellow}},\
  }\bibfield  {title} {\enquote {\bibinfo {title} {Transferability in machine
  learning: from phenomena to black-box attacks using adversarial samples},}\
  }\href {https://arxiv.org/abs/1605.07277} {\bibfield  {journal} {\bibinfo
  {journal} {arXiv:1605.07277}\ } (\bibinfo {year} {2016})}\BibitemShut
  {NoStop}%
\bibitem [{\citenamefont {Tram{\`e}r}\ \emph {et~al.}(2017)\citenamefont
  {Tram{\`e}r}, \citenamefont {Kurakin}, \citenamefont {Papernot},
  \citenamefont {Goodfellow}, \citenamefont {Boneh},\ and\ \citenamefont
  {McDaniel}}]{tramer2017ensemble}%
  \BibitemOpen
  \bibfield  {author} {\bibinfo {author} {\bibfnamefont {F.}~\bibnamefont
  {Tram{\`e}r}}, \bibinfo {author} {\bibfnamefont {A.}~\bibnamefont {Kurakin}},
  \bibinfo {author} {\bibfnamefont {N.}~\bibnamefont {Papernot}}, \bibinfo
  {author} {\bibfnamefont {I.}~\bibnamefont {Goodfellow}}, \bibinfo {author}
  {\bibfnamefont {D.}~\bibnamefont {Boneh}}, \ and\ \bibinfo {author}
  {\bibfnamefont {P.}~\bibnamefont {McDaniel}},\ }\bibfield  {title} {\enquote
  {\bibinfo {title} {Ensemble adversarial training: Attacks and defenses},}\
  }\href {https://arxiv.org/abs/1705.07204} {\bibfield  {journal} {\bibinfo
  {journal} {arXiv:1705.07204}\ } (\bibinfo {year} {2017})}\BibitemShut
  {NoStop}%
\end{thebibliography}%

\clearpage
\setcounter{figure}{0}
\makeatletter
\renewcommand{\thefigure}{S\@arabic\c@figure}
\setcounter{equation}{0} \makeatletter
\renewcommand \theequation{S\@arabic\c@equation}
\renewcommand \thetable{S\@arabic\c@table}

\begin{center} 
{\large \bf Supplementary Material for: Vulnerability of Machine Learning Phases of Matter}
\end{center}

\section{Methods for generating adversarial perturbations} \label{Methods}
In the main text, we have shown that the machine learning approaches to phases of matter based on deep neural networks are extremely vulnerable to adversarial examples: adding a tiny amount of carefully-crafted perturbation, which are imperceptible to human eyes, into the original legitimate  data will cause the phase classifiers to make incorrect predictions with a high confidence level. Here in this section, we give more technical details on how to obtain the adversarial perturbations.

As discussed in the main text, in supervised learning the training data is labeled $\mathcal{D}=\{(\mathbf{x}^{(1)},y^{(1)}),\cdots,(\mathbf{x}^{(n)},y^{(n)})\}$ and the task of obtaining adversarial examples reduces to solving the following optimization problem:
\begin{eqnarray}
\max_{\delta\in\Delta} & \;L(h(\mathbf{x}^{(i)}+\delta;\theta),y^{(i)}). \label{eq:AdvLmaxLoss}
\end{eqnarray}
In the adversarial machine learning literature, a number of methods have been introduced to deal with the above optimization problem. We consider two scenarios in this paper, one is called discrete attack scenario, where the adversarial perturbations are discrete and the original legitimate samples are modified by discrete values; the other is called continuous attack scenario, where the perturbations are continuous and the original legitimate samples are modified continuously. For the discrete attack scenario, we mainly apply the differential evolution algorithm \cite{storn1997differential,das2010differential},  which is a population based optimization algorithm for solving complex multi-modal problems and has recently been used for generating one-pixel adversarial perturbations to fool deep neural networks in image recognition \cite{su2019one}. For the continuous attack scenario,  we use a number of attacking methods, including fast gradient sign method (FGSM) \cite{goodfellow2014explaining,Madry2017Towards}, projected gradient descent (PGD)\cite{Madry2017Towards} and momentum iterative method (MIM) \cite{dong2018boosting}. 

For the case of the ferromagnetic Ising model, we apply both the discrete and continuous attacks, whereas for the case of topological phases of matter we apply only the continuous attacks. We use cleverhans 
 \cite{papernot2018cleverhans} to implement FGSM, PGD, and MIM for both the Ising and chiral topological insulator cases.  In each case, we produce the adversarial samples based on the origin legitimate training set. We define the success ratio as the proportion of adversarial samples that successfully fool the classifier. 
 In the following, we briefly sketch the essential ideas for each attacking methods used in this paper. For each method, we also provide a pseudocode to clearly illustrate how it works.

\subsection{Differential evolution algorithm}\label{Discrete Attack}

Differential evolution is a population based optimization algorithm and  is arguably one of the most powerful stochastic real-parameter optimization algorithms in solving complex multi-modal optimization problems \cite{storn1997differential,das2010differential}. It belongs to the general class of evolutionary algorithms and the computational steps it takes are quite similar to these taken by a standard evolutionary algorithm. 
Yet, unlike traditional evolutionary algorithms, the differential evolution algorithm perturbs the current generation population members with the scaled differences of randomly chosen distinct population members. 
More specifically, during each iteration we randomly generate a new set of candidate solutions (called children) according to the current population (parents), and then compare the children with their corresponding parents, replacing the parents if the children have higher fitness value. 

We apply the differential evolution algorithm in the ``black-box" setting to generate adversarial examples for the ferromagnetic Ising model \cite{su2019one}, where we assume no prior information about the classifier's internal structures and only discrete changes of the samples could be made: we first generate some counterfeit samples by reversing a number of magnetic moments of the legitimate sample randomly. We denote these samples as $X_1,X_2,\dots,X_n$, where $n$ is the population size.  We then feed these samples into the classifier to obtain the confidence probability for each configuration.  \diff{Then we produce new counterfeit samples, which are called children, based on prior samples. Particularly, for each component $X_i'(s)$ of the children $X_i'$, we have the following generation rule:
\begin{eqnarray}
p_s &&\sim U(0,1), \\
X_i'(s)=&&\left\{
\begin{aligned}
&[X_j+M(X_k-X_l)](s), \quad & p_s\leq P \\
&X_i(s), & p_s>P \\
\end{aligned}
\right.
\end{eqnarray}
where $M$ is the mutual factor (larger $M$ leads to larger search radius but take longer time to converge). $j,k,l$ are chosen randomly from $[n]/\{i\}$, $P$ is called the crossover probability. If these children have better performance (i.e., higher confidence probability for the wrong classification category),  then we replace their corresponding parents with these children.  We repeat this procedure with several iterations until it converges and the desired adversarial samples are obtained. For the particular Ising case considered in this paper, we denote every children generation as a sequence of $(a,b,s)^N$, where $(a,b)$ is the flipped spins' positions, $s$ is the spin after the reversing (which is restricted to be either $0$ or $1$ ), and $N$ is the number of changed spins.} A pseudocode representation of the differential evolution algorithm is shown in Algorithm \ref{DEA}.


%

It is worthwhile to mention that the differential evolution algorithm cannot guarantee that the optimal solution will be obtained. It is possible that the algorithm may only yields certain local 
minima. In our scenario, this means that the adversarial examples we obtained may not be the most effective ones to fool the classifier.

\begin{figure}
\begin{algorithm}[H]
\caption{The Differential Evolution Algorithm}
\begin{algorithmic}[1]  
\Require \diff{A legitimate sample $(\vec{x},y)$, the trained model $h(\cdot;\theta)$, the loss function $L$.}
\Require \diff{The iteration number $T$, the population size $n$, the mutual factor $M$, the crossover probability $P$, the number $N$ of spins to flip.}
\Ensure An adversarial example $\vec{x}^*$.
\State Set the position bound $B$ to be the shape of $\vec{x}$
\diff{\State Randomly generate perturbation $X_i=\bigotimes_{j=1}^N(a,b,s)$ with $(a,b)\in B$ and $s=1-\vec{x}(a,b)$ for $i=1,2,\dots,n$
\State Get adversarial sample $\vec{x}^*_i$ by adding $X_i$ on $\vec{x}$ for $i=1,2,\dots,n$}
\For {$t=1,2,\dots,T$}
\For {$i=1,2,\dots,n$}
\diff{\State Randomly pick distinct $j,k,l\in [n]/\{i\}$ 
\State Generate children: $X_i'=X_j+M(X_k-X_l)$
\For {$s=1,2,\dots,N$}
\State Randomly pick $p$ from $U(0,1)$
\If {$p>P$}
    \State $X_i'(s) = X_i(s)$
\EndIf
\EndFor
\State Get adversarial sample $\vec{x}^{*'}_i$ by $X_i'$
\If {$L(h(\vec{x}^{*'}_i;\theta),y)>L(h(\vec{x}^{*}_i;\theta),y))$}
        \State $X_i=X_i'$
        \State $\vec{x}^{*}_i=\vec{x}^{*'}_i$
\EndIf}
\EndFor
\EndFor
\State Find the $X_p$ among $\{X_1, X_2, \dots, X_n\}$ that has the highest confidence probability for the wrong classification category, apply $X_p$ to $\vec{x}$ to get $\vec{x}^*$\\
\Return $\vec{x}^*$
\end{algorithmic}  \label{DEA}
\end{algorithm}
\end{figure}


\subsection{Fast gradient sign method}
The fast gradient sign method is a simple one-step scheme for solving Eq. (\ref{eq:AdvLmaxLoss})  and has been widely used in the adversarial machine learning community \cite{goodfellow2014explaining,Madry2017Towards}.  Before introducing this method, let us first introduce the fast gradient method (FGM).

We work in a white-box attack setting, where full information about the classifier is assumed. Our goal is to maximize the loss function for a particular input data $\vec{x}$ to generate the adversarial sample $\vec{x}^*$. Since we know all parameters of the model, we can compute the fastest increasing direction on the position $\vec{x}$, which is just the gradient of the loss function: $\nabla_x L(h(\vec{x};\theta),y)$. The FGM is a one-step attack which perturbs $\vec{x}$ along the direction of the gradient with one particular stepsize:
\begin{eqnarray}
\delta_{\text{FGM}}=\max_{\delta\in \Delta}\langle\nabla_x L(h(\vec{x};\theta),y),\delta\rangle.
\end{eqnarray}
The perturbation is constrained within $l_p$-norm bound:  \(\|\delta\|_{p}\leq \epsilon\).

If we take $l_{\infty}$-norm bound, we get a simple rule for obtaining the adversarial perturbation via FGSM:
\begin{eqnarray}
\delta_{\text{FGSM}} = \epsilon\cdot \text{sign}(\nabla_x L(h(\vec{x};\theta),y)). \label{FGSMupdate}
\end{eqnarray}
For different problems, there are other particular perturbation bounds as well. One of the most useful bounds is the rectangular-box-like bound, where each component of the adversarial sample is bounded by some constant numbers $x_{\text{min}} \leq x \leq x_{\text{max}}$. For example, for the case of chiral topological insulators studied in this paper, we require that every component of $\vec{x}$ be bounded by $[-1,1]$. If the adversarial sample has components exceeding this bound, FGSM simply change the value of this component to be the value of either $x_{\text{min}}$ or $x_{\text{max}}$.  A pseudocode representation of the fast gradient sign method is shown in Algorithm \ref{FGSM}.

\begin{figure}[htp]
\begin{algorithm}[H]
\caption{Fast Gradient Sign Method}
\begin{algorithmic}[1]  

\Require \diff{The trained model $h(\cdot, \theta)$, the loss function $L$, the legitimate sample $(\vec{x},y)$.}
\Require The perturbation strength $\epsilon$, upper and lower bound $x_{\text{min}}, x_{\text{max}}$.
\Ensure An adversarial example $\vec{x}^*$.
\State Input $\vec{x}$ into $h$ to get $\nabla_x L(h(\vec{x};\theta),y)$
\For {each component $\vec{x}(i)$ of $\vec{x}$}
	\State $\delta_i=\epsilon\cdot \text{sign}(\nabla_x L(h(\vec{x};\theta),y)(i))$
	\State $\vec{x}^*(i)=\vec{x}(i)+\delta_i$
	\If {$\vec{x}^*(i)>x_{\text{max}}$}
		\State $\vec{x}^*(i)=x_{\text{max}}$
	\EndIf
	\If {$\vec{x}^*(i)<x_{\text{min}}$}
		\State $\vec{x}^*(i)=x_{\text{min}}$
	\EndIf
\EndFor \\
\Return $\vec{x}^*$
\end{algorithmic}  \label{FGSM}
\end{algorithm}
\end{figure}

\subsection{Projected gradient descent method}\label{pgd}

As shown in Eq. (\ref{FGSMupdate}), one may interpret the FGSM as a simple one-step scheme for maximizing the inner part of the saddle point formulation. 
With a small stepsize, FGSM may perform well. But with a large stepsize, FGSM can perform poorly since the gradient of the loss function may change significantly during this step. To deal with this problem, a more powerful method is its multi-step variant, which is called the projected gradient descent method (PGD). The basic idea of PGD is to use FGSM methods with multiple times ($T$) and perform projections iteratively to enforce that the perturbation is within an appropriate region \cite{Madry2017Towards}. At each step, we check if the proposed update has moved out of the region, and apply a projection back if it does. So the rule for updating is
\begin{eqnarray}
\vec{x}_{t+1}=\pi_{C}(\vec{x}_t+\alpha\cdot \text{sign}(\nabla_x L(\theta,\vec{x}_t,y))),
\end{eqnarray}
where $\alpha=\frac{\epsilon}{T}$ is the stepsize and $\pi_{C}$ is the projection operation which projects those points out of the chosen appropriate region [denoted as $\Delta$ in Eq. (\ref{eq:AdvLmaxLoss})]
 back. In our scenarios, the permitted region we choose  is the region that restricts every component of $\vec{x}$ to be in $[x_{\text{min}},x_{\text{max}}]$, therefore, $\pi_C$ is simply the projection for each component into $[x_{\text{min}},x_{\text{max}}]$. A pseudocode representation for the projected gradient descent method is shown in Algorithm \ref{PGDM}.

\begin{figure}[htp]
\begin{algorithm}[H]
\caption{Projected Gradient Descent Method}
\begin{algorithmic}[1]  

\Require \diff{The trained model $h(\cdot, \theta)$, loss function $L$, the legitimate sample $(\vec{x},y)$.}
\Require The perturbation strength $\epsilon$, iteration number $T$, upper and lower bound $x_{\text{min}}, x_{\text{max}}$.
\Ensure An adversarial example $\vec{x}^*$.
\State $\vec{x}_0=\vec{x}$
\State  $\alpha=\frac{\epsilon}{T}$
\For {$i=1,\dots,T$}
	\State Input $\vec{x}_{i-1}$ into $h$ to get $\nabla_x L(h(\vec{x}_{i-1};\theta),y)$
	\For {every component $\vec{x}_{i-1}(j)$ of $\vec{x}_{i-1}$}
		\State $\delta_j=\alpha\cdot \text{sign}(\nabla_x L(h(\vec{x}_{i-1};\theta),y))(j)$
		\State $\vec{x}_{i}(j)=\vec{x}_{i-1}(j)+\delta_j$
		\If {$\vec{x}_{i}(j)>x_{\text{max}}$}
			\State $\vec{x}_{i}(j)=\pi_C(\vec{x}_{i}(j))=2x_{\text{max}}-\vec{x}_{i}(j)$
		\EndIf
		\If {$\vec{x}_{i}(j)<x_{\text{min}}$}
			\State $\vec{x}_{i}(j)=\pi_C(\vec{x}_{i}(j))=2x_{\text{min}}-\vec{x}_{i}(j)$
		\EndIf
	\EndFor
\EndFor \\
\Return $\vec{x}^*=\vec{x}_T$
\end{algorithmic}  \label{PGDM}
\end{algorithm}
\end{figure}

\subsection{Momentum iterative method}\label{momentum-iterative-method-mim}
The FGSM assumes the sign of the gradient of loss function will not change around the data point and generates an adversarial example by applying the sign of the gradient to a legitimate example only once. However, in many practical applications the assumption may not hold when the distortion is large, rendering the adversarial example generated by FGSM ``under-fits" the model. On the other hand, iterative FGSM like PGD moves the counterfeit examples gradually in the direction of the sign of the gradient in each iteration and hence can easily drop into poor local 
extremums and ``overfit" the model.   To deal with such a dilemma, one can integrate momentum into the iterative FGSM so as to stabilize update directions and escape from local extremums \cite{dong2018boosting}. This is the essential idea of the momentum iterative method. 


For a $T$ iterations attack with $l_{\infty}$-norm constraint $\epsilon$, in every iteration we calculate the gradient descent direction and add the gradient descent direction in the last iteration with a decay factor $\mu$ as the accelerated velocity:
\begin{eqnarray}
a_{t+1}=\mu\cdot a_t + \frac{\nabla_{x_t} L(h(\vec{x}_t;\theta),y)}{||\nabla_{x_t} L(h(\vec{x}_t;\theta),y)||},
\end{eqnarray}
and the rule for updating is:
\begin{eqnarray}
x_{t+1}=x_t+\alpha\cdot \text{sign}(a_{t+1}),
\end{eqnarray}
where $\alpha=\frac{\epsilon}{T}$ is the stepsize. A pseudocode representation for the momentum iterative method is shown in Algorithm \ref{MIM}.

\begin{figure}
\begin{algorithm}[H]
\caption{Momentum Iterative Method}
\begin{algorithmic}[1]  
\Require \diff{The trained model $h(\cdot, \theta)$, loss function $L$, the legitimate sample $(\vec{x},y)$.}
\Require The perturbation strength $\epsilon$, iteration number $T$, decay factor $\mu$, upper and lower bound $x_{\text{min}}, x_{\text{max}}$.
\Ensure An adversarial example $\vec{x}^*$.
\State $\vec{x}_0=\vec{x}$, $a_0=0$
\State $\alpha=\frac{\epsilon}{T}$
\For {$t=1,\dots,T$}
	\State Input $\vec{x}_{i-1}$ into $h$ to get $\nabla_x L(h(\vec{x}_{t-1};\theta),y)$
	\State $a_{t}=\mu\cdot a_{t-1} + \frac{\nabla_x L(h(\vec{x}_{t-1};\theta),y)}{||\nabla_x L(h(\vec{x}_{t-1};\theta),y)||}$
	\For {each component $\vec{x}_{t-1}(j)$ of $\vec{x}_{t-1}$}
		\State $\delta_j=\alpha\cdot \text{sign}(a_t)(j)$
		\State $\vec{x}_{t}(j)=\vec{x}_{t-1}(j)+\delta_j$
		\If {$\vec{x}_{t}(j)>x_{\text{max}}$}
			\State $\vec{x}_{t}(j)=\pi_C(\vec{x}_{t}(j))=2x_{\text{max}}-\vec{x}_{t}(j)$
		\EndIf
		\If {$\vec{x}_{t}(j)<x_{\text{min}}$}
			\State $\vec{x}_{t}(j)=\pi_C(\vec{x}_{t}(j))=2x_{\min}-\vec{x}_{t}(j)$
		\EndIf
	\EndFor
\EndFor \\
\Return $\vec{x}^*=\vec{x}_T$
\end{algorithmic}  \label{MIM}
\end{algorithm}
\end{figure}

\begin{figure}
\centering
\includegraphics[width=0.47\textwidth]{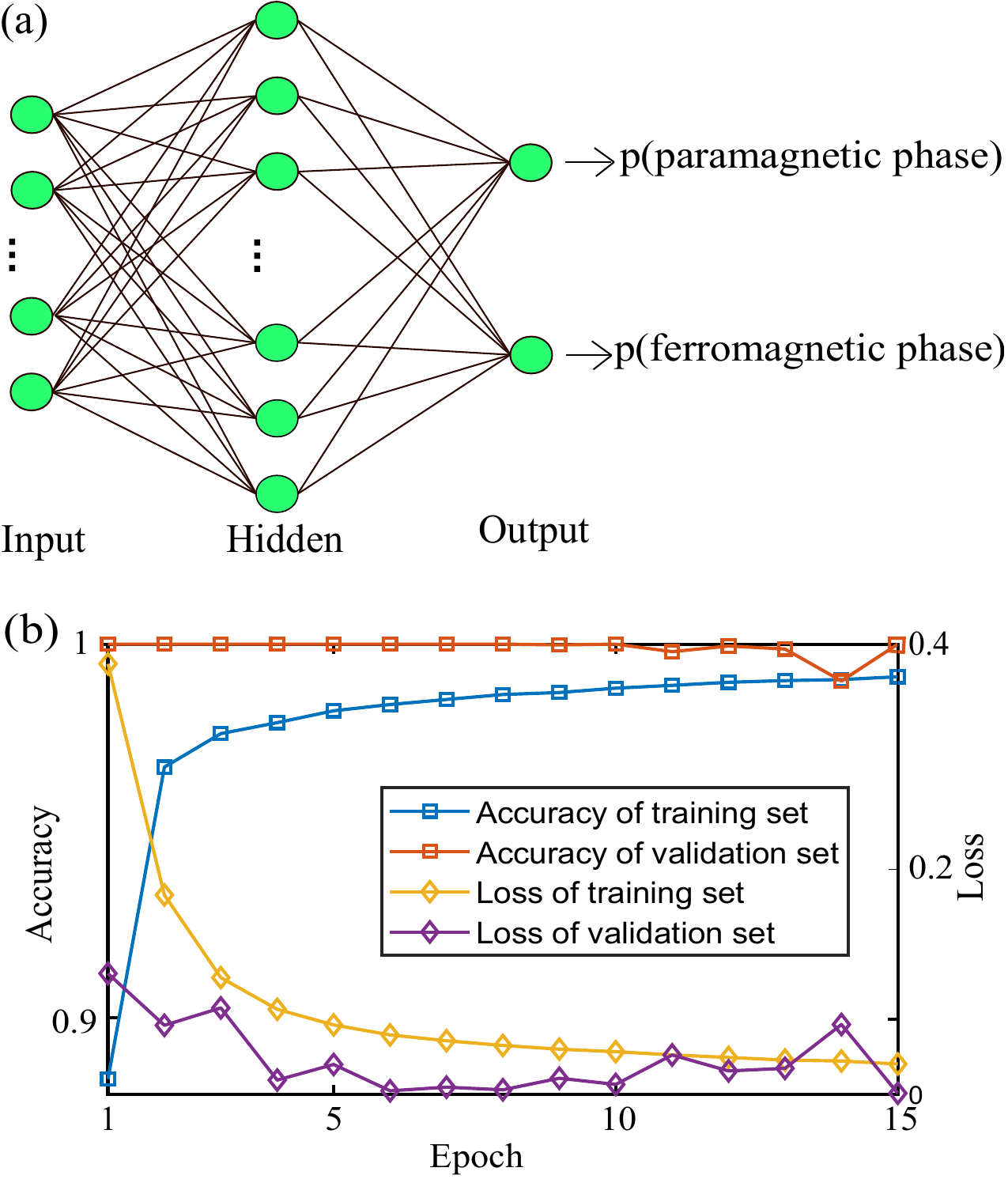}
\caption{Machine learning ferromagnetic/paramagnetic phases of the Ising model. (a) The classifier is a fully connected feed forward neural network.   It consists of one input layer with $900$ neurons which have one-to-one correspondence to the spins of the Ising model, one hidden layer with $100$ sigmoid neurons,  and one output layer with two softmax neurons outputting the probabilities of the paramagnetic and ferromagnetic phases. (b) The training process. The classifier is trained with numerically simulated data at $40$ different temperatures from  $T=0$ to $T=3.54$.  The training set contains $90000$ samples, each sample is a array with length $900$. The validation set is of size $10000$ and the test set is of size $10250$. We use RMSprop optimizer with batch size of $256$ and learning rate of $10^{-3}$. The accuracy is the correct classification percentage and the loss is the value of cross-entropy. 
\label{fig:ising_train}}
\end{figure}

\section{More details on the two  concrete examples}
\diff{In this section, we provide more technical details on the neural network structures of the classifiers, the training process, and the analysis of adversarial examples} In addition, we provide more numerical simulation results for both the examples of the Ising model and the chiral topological insulator.

\begin{figure}
\centering
\includegraphics[width=0.47\textwidth]{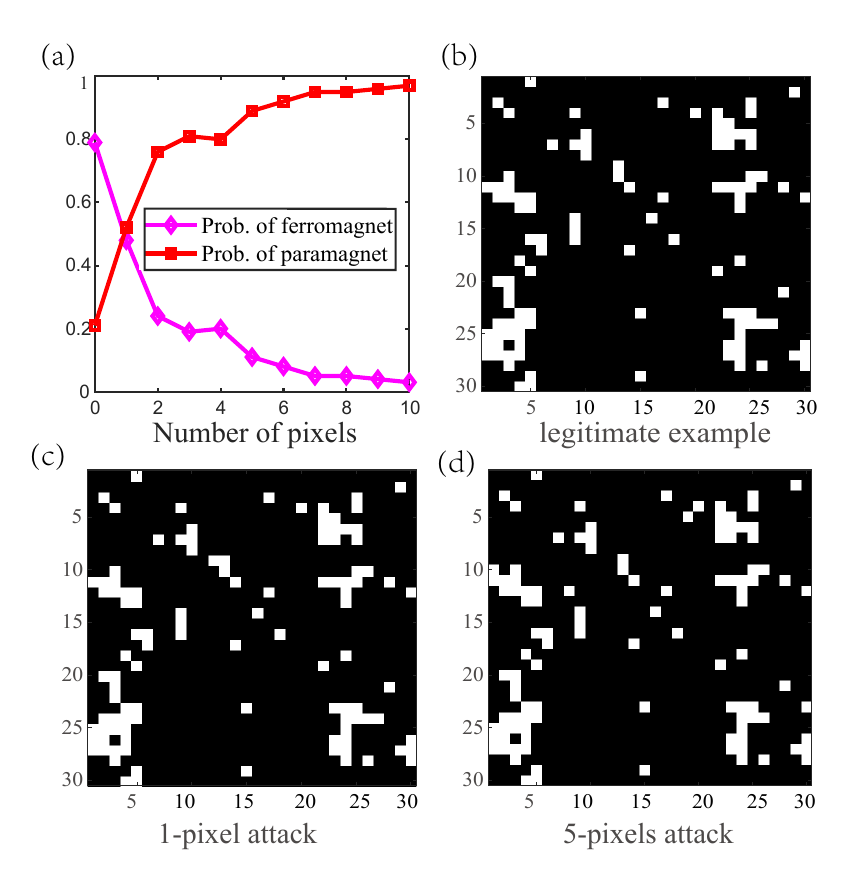}
\caption{Performance of the differential evolution algorithm for the case of ferromagnetic Ising model.  In this figure, the population $n$ is set to be $100$ and the mutual factor $M$ set to be $30$. The algorithm stops and returns the adversarial samples when the confidence probability converges with increasing number of  iterations. (a) The confidence probabilities for the ferromagnetic and paramagnetic categories versus the number of pixels that are allowed to be changed (i.e., spins allowed to be flipped).  We randomly choose a legitimate sample (here the $5067^{th}$ sample in the validation set) which is correctly classified by the neural network to  belong to the ferromagnetic phase with confidence $80\%$. 
(b) The legitimate sample from the ferromagnetic phase. Here, each small square corresponds to a spin and black (white) color means the corresponding spin points down (up).   (c) An adversarial sample with one pixel changed. The classifier misclassifies this modified sample into the paramagnetic category with confidence  $52\%$. (d) An adversarial sample with five pixels changed. The classifier misclassifies this modified sample into the paramagnetic phase with confidence  $90\%$.
\label{fig:deg}}
\end{figure}

\subsection{The ferromagnetic Ising model}
In the main text, we have shown that adding a tiny amount of adversarial perturbation as small as a single pixel can lead the classifier to misclassify a spin-configuration image from the ferromagnetic phase  into the paramagnetic category. In this example, our phase classifier is  a fully connected feed-forward neural network, which is composed of an input layer with $900$ neurons,  a hidden layer with $100$ sigmoid neurons, and an analogous output layer with two sigmoid neurons,  as shown in Fig. \ref{fig:ising_train}(a). The input data is the equilibrium spin configurations sampled from Monte Carlo simulations, same as in Ref.  \cite{Carrasquilla2017Machine}. We use  $0$ and $1$ to represent whether the spin is up or down. The lattice size is fixed to be $30\times30$, and therefore the input data $\vec{x}$  are $\{0,1\}$ arrays with length $900$. The training and validation sets are both numerically generated with Monte Carlo simulations \cite{Carrasquilla2017Machine}, and their sizes are $90000$ and $10000$, respectively.  
We use the RMSprop as the optimizer with batch size of $256$ and the learning rate is set to be $10^{-3}$. In Fig. \ref{fig:ising_train}(b), we plot the results for the training process. From this figure, it is clear that the accuracy increases (the loss decreases) as  the number of epochs increases, and after $15$ epochs  the network can successfully classify samples from the validation/test set with a high accuracy larger than $97\%$.

\begin{figure*}
\centering
\includegraphics[width=0.9500\textwidth]{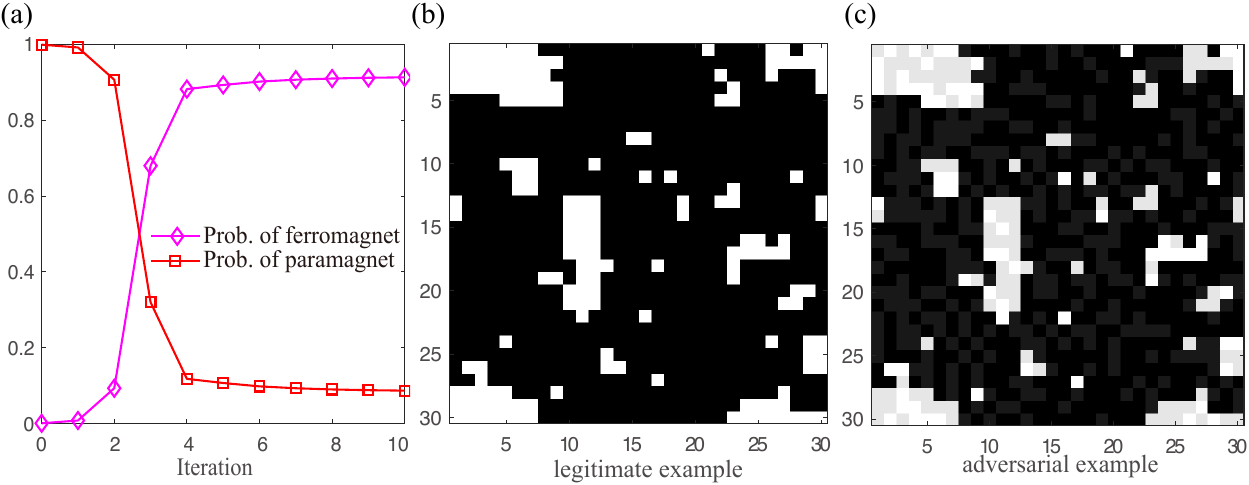}
\caption{Performance of the momentum iterative method (MIM) for the case of Ising model. (a) The classification probabilities of the ferromagnetic and paramagnetic phases as a function of the iteration number.It is clear that  after around three iterations, the classifier will begin to misclassify the samples, and after ten iterations the slightly modified samples will be identified as belongs to the ferromagnetic category with confidence $\geq 90\%$. (b) A randomly chosen legitimate sample from the paramagnetic phase. (c) An adversarial example obtained by MIM, which is slightly different from the original sample. Here, the perturbation is restricted to be within $\|\delta\|_{\infty}\leq 0.1$.
\label{fig:continuous attack}}
\end{figure*}

After the training process, we fix the parameters of the classifier and utilize different methods to generate adversarial examples. The first method we use is the differential evolution algorithm, which is a discrete attack method. Fig. 2(a) of the main text gives an adversarial example that only differs with the original legitimate one by a single pixel. Intuitively, if we modify the original sample by flipping more spins, the confidence probability for the classifier to misclassify the modified sample will increase. This is also verified in our numerical simulations and partial of our results are shown in Fig. \ref{fig:deg}. In Fig. \ref{fig:deg}(a), we randomly choose a legitimate sample from the ferromagnetic phase, which is shown in Fig. \ref{fig:deg}(b). Without changing any pixel (flipping a spin), the classifier will correctly identify the sample as from the ferromagnetic phase with a confidence level $\approx 80\%$. However, this confidence probability will decrease rapidly as the number of pixels that are allowed to change increases. This is clearly demonstrated in \ref{fig:deg}(a). Fig. \ref{fig:deg}(c) and Fig. \ref{fig:deg}(d) plot two adversarial examples with one and five pixels of the original sample (Fig. \ref{fig:deg}(b)) changed, respectively. For this particular legitimate sample, changing one pixel (five pixels) will lead the classifier to misclassify it with confidence $52\%$ ($90\%$). We note that in order to obtain Fig.2 (b) and Fig. 2 (c) of the main text, the hyper parameters we used are the same as these in Fig. \ref{fig:deg}.

\begin{table}
\setlength{\tabcolsep}{3mm}
\diff{\begin{tabular}{cccc}
\toprule
Position & Original spin & Value in AM&  $P'$(ferro)\\
\midrule
18 & $ - $ & $ -2.751 $ & $ 0.479 $ \\
95 & $ - $ & $ -2.606 $ & $ 0.464 $ \\
108 & $ - $ & $ -2.736 $ & $ 0.462 $ \\
116 & $ - $ & $ -2.651 $ & $ 0.440 $ \\
229 & $ - $ & $ -3.503 $ & $ 0.402 $ \\
251 & $ + $ & $ -3.554 $ & $ 0.902 $ \\
289 & $ - $ & $ -2.727 $ & $ 0.461 $ \\
337 & $ - $ & $ -3.296 $ & $ 0.446 $ \\
356 & $ - $ & $ -2.867 $ & $ 0.451 $ \\
396 & $ - $ & $ -2.830 $ & $ 0.461 $ \\
445 & $ - $ & $ -2.931 $ & $ 0.424 $ \\
521 & $ - $ & $ 2.671 $ & $ 0.881 $ \\
534 & $ - $ & $ -3.277 $ & $ 0.396 $ \\
561 & $ - $ & $ -2.610 $ & $ 0.481 $ \\
768 & $ + $ & $ -2.767 $ & $ 0.880 $ \\
783 & $ - $ & $ -2.637 $ & $ 0.445 $ \\
829 & $ + $ & $ -2.765 $ & $ 0.892 $ \\
\bottomrule
\end{tabular}}
\caption{\diff{The result of single spin flips. The original spin configuration is classified as ferromagnetic phase with confidence $P_{\text{ferro}}=72\%$. The ``position" is the spin index on the $30\times30$ lattice. The ``original spin" represents the original spin direction in the legitimate example. The ``value in AM" represents the position's corresponding value in the activation map. ``$P'$(ferro)" means the classifier's confidence to identify the sample into ferromagnetic phase after the a single spin flip on the original legitimate example. From the table, we can find that spin changes to have the different sign to the corresponding value in AM will dramatically decrease the condifence of being ferromagnetic phase.} \label{table: spin flips}}
\end{table}

As discussed in the main text, we may also regard $H_{\text{Ising}}$ as a quantum Hamiltonian and the input data to be the local magnetization, and thus we allow the input data to be continuously modified. In this case, we can use different methods, such as FGSM, PGD, and MIM discussed in Sec. \ref{Methods}, to generate adversarial examples.  Partial of our results are shown in Fig. \ref{fig:continuous attack}. In Fig. \ref{fig:continuous attack}(a), we randomly choose a sample from the paramagnetic phase, which is plotted in Fig.  \ref{fig:continuous attack}(b). At the beginning, the classifier can correctly identify this sample as in the paramagnetic category with confidence $\geq 99\%$. We then use MIM to modify the original sample and after around three iterations, the classifier will begin to make incorrect predictions, and after ten iterations it will misclassify the sample to be in the ferromagnetic phase with confidence $\geq 90\%$. Fig. \ref{fig:continuous attack}(c) shows the corresponding adversarial example obtained by MIM after ten iterations.

\begin{figure*}
\centering
\includegraphics[width=0.9500\textwidth]{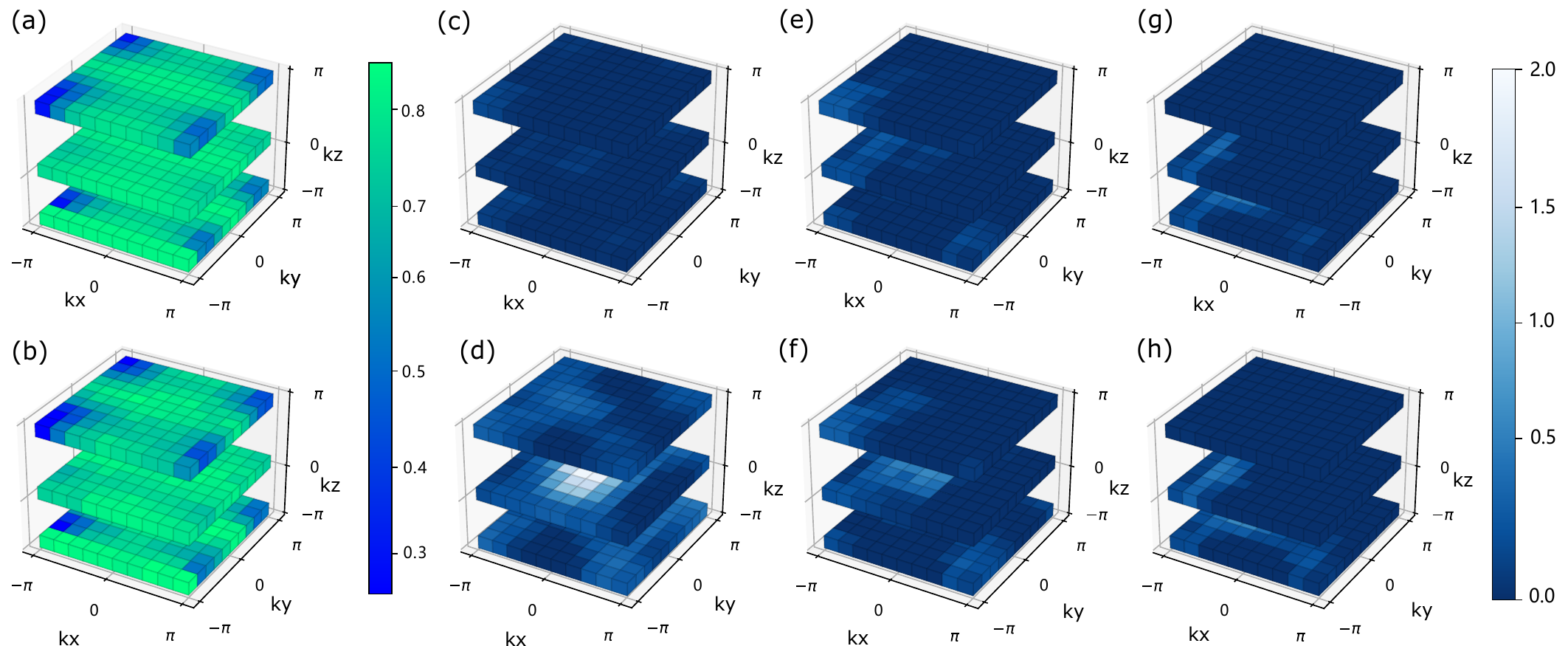}
\caption{\diff{(a) A legitimate sample of the first component of the input data (which is related to the density matrix in the momentum space).
Here, only slices corresponding to $-\pi, 0, 4\pi/5$ are displayed. (b) An adversarial example obtained by FGM, which only differs with
the sample in (a) by a tiny perturbation and has average fidelity 0.997. (c)(d)(e)(f) The activation maps (AM) of the sixth kernel in the first
convolutional layer under different settings. (c) The average AM on all samples in the test set with $\chi^{(m)}=0$. (d) The average AM on $\chi^{(m)}=1$. (e) The AM obtained by taking the legitimate sample (a) as the input to the original classifier and (f) is taking the adversarial
example (b) as input. (g)(h) are similar to (e)(f) but change the original classifier into the classifier after adversarial training.}
\label{fig:attmap_Chiral}}
\end{figure*}

\begin{figure}
\centering
\includegraphics[width=0.47\textwidth]{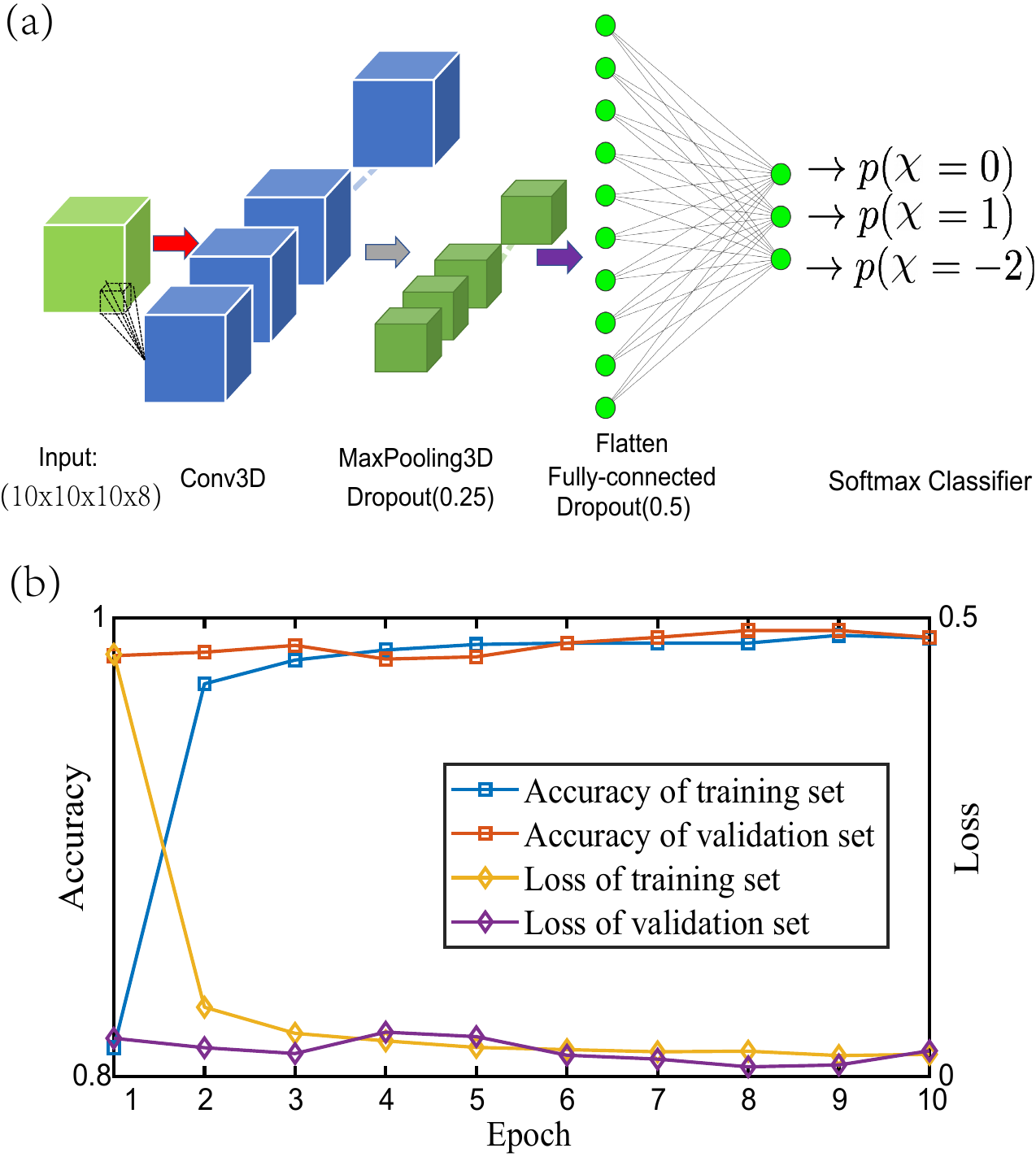}
\caption{Learning topological phases with a 3D convolutional neural network (CNN).  (a) The structure of the CNN phase classifier. (b) The training process. Here, the loss function is chosen to be the cross-entropy. It is clear that  after ten epochs, the classifier can successfully identify the samples from both the training and validation sets with accuracy $\approx 99\%$.   
\label{fig:chiral_train}}
\end{figure}

\diff{In Fig.~2(a) of the main text, we briefly introduced the idea of the classifier's activation map on the Ising model, now we explain in detail about how we get this activation map. Since the classifier structure we used for the Ising model is rather simple, the inference of the classifier can be strictly written as the following:}
\begin{eqnarray}
\diff{
\begin{aligned}
P&=\text{softmax}(W_2\cdot \text{sigmoid}(W_1\cdot x+b_1)+b_2)\\
&\sim (W_2\cdot W_1)x
\end{aligned}}
\end{eqnarray}
\diff{where $x$ is the input Ising configuration with shape $900\times1$, $W_1$ is the first layer weight with shape $100\times900$, and $b_1$ is the first layer bias with shape $100\times1$. Similarly, $W_2$ is the second layer weight with shape $2\times100$ and $b_2$ is the second layer bias with shape $2\times1$. We can see that $P_1$, which is the first element of final prediction $P$ and represents the confidence of being ferromagnetic phase, is proportional to $(W_2\cdot W_1\cdot x)_1$. Therefore, we denote $(W_2\cdot W_1)_{1, :}$, which is the fist row of $W_2\cdot W_1$, as the activation map $W_{\text{ferro}}$.  $W_{\text{ferro}}$ has shape $1\times 900$ and approximately represents $\partial P_{\text{ferro}}/\partial x$, denoting the weight of each spin on final prediction to the ferromagnetic phase. The spin with the same sign as the corresponding position's value in activation map will contributes to being classified as the ferromagnetic phase and vice versa. This explains why a single spin flip can change the classifier's prediction dramatically, either increase or decrease. Using the legitimate example shown in main text Fig.~2(a), we try single spin flips on all positions in the activation map with value larger than $2.6$ and list the result in Table \ref{table: spin flips}. We can find that single spin flips which make their signs change to the opposite of values in the AM  can make the classifier incorrectly identifies the example into paramagnetic phase (i.e., $P'$(ferro)$<0.5$). This result, combined with the highly non-uniform activation map, show that the classifier’s prediction relies heavily on only several particular spins. This does not agree with the physically defined order parameter $M=\frac{|\sum_i \sigma_i|}{N}$ where each spin contributes equally, indicating the model does not fully capture the underlying physical principles.}

\subsection{Topological phases of matter}\label{ChiralTI}

For the example of topological phases of matter, the classifier we consider is a 3D convolutional neural network (CNN), as shown in Fig.~\ref{fig:chiral_train}(a). It consists of two 3D convolution layers, a 3D max pooling layer, a dropout layer with rate $0.4$ to avoid overfitting, and a flattening layer connected with two fully-connected layers with $0.55$ dropout. The output layer is a softmax layer outputting the probability for the three possible topological phases. We use the RMSprop as the optimizer with batch size of $128$. The loss function is chosen to be the cross-entropy. The learning rate is set to be $10^{-3}$.

In our scenario, the input data are the density matrices on a $10\times10\times10$ momentum grid and we express each density matrix $\rho$ as  \cite{Lian2019Machine}:
\begin{eqnarray}
\rho=\frac{1}{3}(\mathbb{I}+\sqrt{3}\mathbf{b}\cdot\vec{\lambda}),\label{density}
\end{eqnarray}
where $\vec{\lambda}$ is a vector consists of the eight Gell-Mann matrices, $\mathbb{I}$ is the three-by-three identity matrix, and $\mathbf{b}=(b_1,b_2,\cdots, b_8)$ with $b_i=\frac{1}{2}\sqrt{3}\text{tr}(\rho\lambda_i)$. Therefore, in this representation of the density matrices each sample of the input data has the form $10\times10\times10\times8$, which can be regarded as $10 \times 10 \times 10$ pixels image with $8$ color channels. To train the network, we numerically generate $5001$ samples as the training set and $2001$ samples as the validation set with parameter $h$ varied uniformly from $-5$ to $5$. The training process is shown in Fig.~\ref{fig:chiral_train}(b). The performance of the training process is shown in Fig.~\ref{fig:chiral_train}(b). From this figure, the training accuracy for the training set increases rapidly at the beginning and then saturates at a high value ($\approx 99\%$), whereas the loss for the training set decrease rapidly at the beginning and then saturate at a small value ($\approx 0.05$). This indicates that the classifier performs remarkably well on the legitimate samples.

After the training was done, we use three different methods, namely FGSM, PGD, and MIM, to generate adversarial examples. We find that all these methods work notably well and can generate adversarial examples with success ratio larger than $76\%$ (i.e., for more than $76\%$ of the legitimate samples, these methods can successfully output the corresponding adversarial examples) with the perturbation bounded by $||\delta||_{\infty} \leq 0.2$. In order to obtain Fig.~3(b) of the main text, we randomly choose a sample  from the category with topological invariant $\chi^{(m)}=1$. At the beginning, the classifier can successfully identify this sample with almost unit confidence probability  ($\approx 99.5\%$). Here, we use MIM to generate adversarial perturbations with restriction $||\delta||_{\infty}\leq 0.05$. The classifier will begin to misclassify the slightly modified sample to be in the category $\chi^{(m)}=0$ after about three iterations. The confidence probability for the misclassification approaches $98\%$ after four iterations and begins to saturate at this value. \diff{The original legitimate sample is shown in Fig. \ref{fig:attmap_Chiral}(a) and Fig. \ref{fig:attmap_Chiral}(b) plots its corresponding adversarial example obtained by MIM. As discussed in Eq. (\ref{density}), each density matrix is represented by a vector $\mathbf{b}$ of length eight. For ease of visualization,  in Fig. \ref{fig:attmap_Chiral}(a-b) we plot only the first component of $\mathbf{b}$, namely $b_1$,  for each momentum point. } We mention that one can also use the experimental data obtained recently in Ref. \cite{Lian2019Machine} with a solid-state simulator to generate adversarial examples. This is also observed in our our numerical simulations. 

\diff{To demonstrate why this adversarial example can mislead the powerful topological phase classifier, we study the activation map of each kernel in the first convolutional layer. We find that the sixth kernel has totally different activation patterns on topologically trivial and nontrivial phases, implying the classifier uses this kernel as a strong indicator for different phases. To verify this statement in depth, we calculate the correlations between the sixth kernel's activation maps on $\chi^{(m)}=0,1,-2$ phases. The metric we use is Normalized Cross Correlation (NCC) $\rho$, which is expressed as the following:
\begin{eqnarray}
\diff{\rho(I_1, I_2)=\frac{\sigma(I_1, I_2)}{\sqrt{D(I_1)\cdot D(I_2)}}}
\end{eqnarray}
where $I_1, I_2$ are two activation maps, $D(\cdot)$ is the variance and $\sigma(\cdot,\cdot)$ is the covariance. It is easy to verify that $|\rho|\leq 1$ and larger $\rho$ indicates stronger correlation. The result is shown in Table \ref{table: NCC}. We can find that the activation map of topologically nontrivial phases ($\chi^{(m)}=1,-2$) are highly correlated ($\rho=0.971$), but have little correlation with topologically trivial phase ($\rho=0.110, 0.244$ respectively).}

\diff{Then the functionality of adversarial example can be explained by the increase of correlation with the activation maps of incorrect phases. In Fig. 2(c) of the main text we show one layer of the activation map of the sixth kernel, we now show the layer with $k_z=-\pi, 0, 4\pi/5$ in Fig. \ref{fig:attmap_Chiral}(c-f), corresponding to the layers in Fig. \ref{fig:attmap_Chiral}(a-b). We can clearly see that the activation map of adversarial example ($I_{\text{adv}}$) becomes more correlated with average activation map of $\chi=1$ ($I_{\chi=1}$). This can also be verified from Table \ref{table: NCC}, the NCC with $I_{\chi=1}$ changes from $0.235$ to $0.689$. This means that the sixth kernel's behaviour on the adversarial example is more similar as the $\chi=1$ phase's, which makes the classifier do the incorrect prediction.}

\begin{table}
\setlength{\tabcolsep}{2mm}
\diff{
\begin{tabular}{c|ccc}
\toprule
Before Adv. Training &  $I_{\chi=0}$ & $I_{\chi=1}$ &  $I_{\chi=-2}$ \\
\midrule
$I_{\chi=0}$ &1 & 0.244 & 0.110 \\
$I_{\chi=1}$ & 0.244 & 1 & 0.971 \\
$I_{\chi=-2}$ & 0.110 & 0.971 & 1\\
\midrule
$I_{\text{leg}}$ & 0.566 & 0.235 & 0.166 \\
$I_{\text{adv}}$ & 0.392 & 0.689 & 0.641 \\
\toprule
After Adv. Training &  $I_{\chi=0}$ & $I_{\chi=1}$ &  $I_{\chi=-2}$ \\
\midrule
$I_{\chi=0}$ &1 & 0.150 & -0.147 \\
$I_{\chi=1}$ & 0.150 & 1 & 0.789 \\
$I_{\chi=-2}$ & -0.147 & 0.789 & 1\\
\midrule
$I_{\text{leg}}$ & 0.776 & 0.158 & -0.103 \\
$I_{\text{adv}}$ & 0.661 & 0.231 & -0.038 \\
\bottomrule
\end{tabular}}
\caption{\diff{The normalized cross correlations between different kinds of activation maps of the sixth kernel before and after adversarial training. $I_{\chi=i}$ represents the average activation map for ${\chi=i}$ phase, where $\chi=0$ is topologically trivial and $\chi=1,-2$ are topologically nontrivial. $I_{\text{leg}}$ is the activation map of the legitimate example with $\chi=0$ and $I_{\text{adv}}$ is the activation map of the adversarial example.}\label{table: NCC}}
\end{table}

\begin{figure}
\centering
\includegraphics[width=0.5\textwidth]{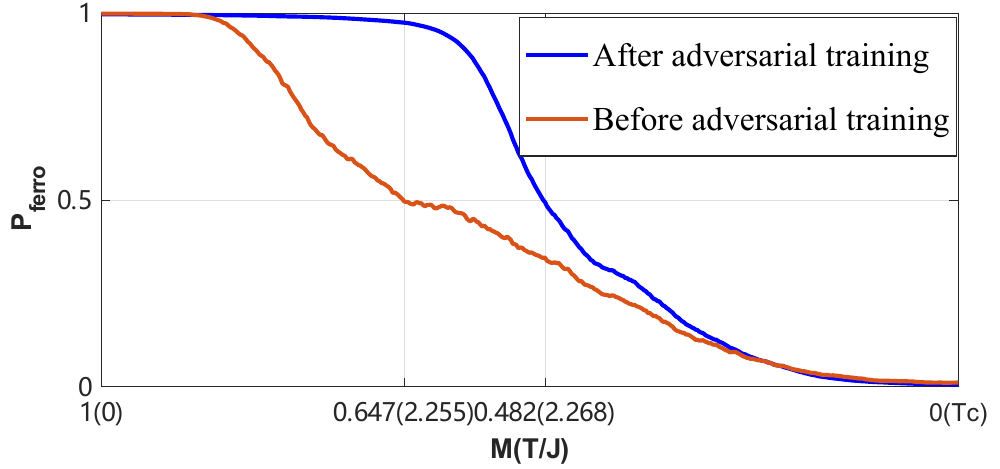}
\caption{\diff{The 2D result of Fig.~4.(c-d) of the main text. The plot is obtained by taking the average on configurations with the same $M$. After adversarial training, the classifier's prediction clip gets sharper and closer to the theoretical transition point ($T_c=2.269$).}
\label{fig: transition_averaging}}
\end{figure}

\begin{figure*}
\centering
\includegraphics[width=0.9\textwidth]{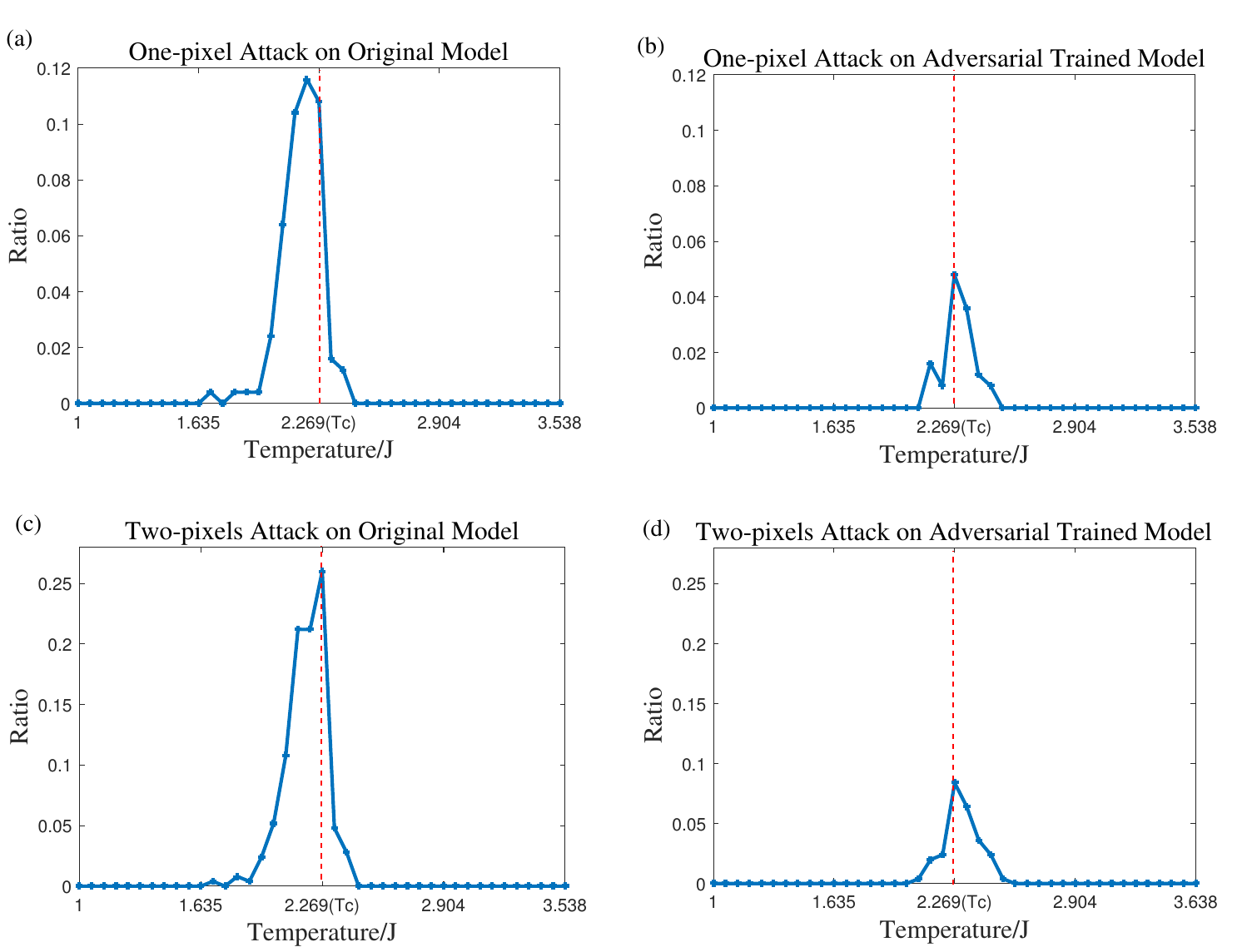}
\caption{\diff{One pixel and two pixels attacks' results on the original classifier and refined classifier after adversarial training against FGSM attacks. (a)(b) show the results of the one pixel attack. It is clear that the model is more vulnerable near critical temperatures $T_c$. After adversarial
training, adversarial samples ratio decreases a lot and the peak moves to $T_c$ exactly, which means the
model becomes more robust. (c)(d) show the results of the two pixels attack and a similar conclusion can be drawn.}
\label{fig: onePixel_against_FGSM}}
\end{figure*}

\section{Adversarial training}\label{defense}
In order to increase the robustness of the deep neural networks to adversarial perturbations, a number of methods have been developed in the adversarial machine learning literature. The simplest and most straightforward one is adversarial training \cite{Madry2017Towards}. Its essential idea is to first generate a substantial amount of adversarial examples with certain attacking methods and then retrain the classifier with both the original legitimate data and the crafted data. After retraining, the classifier will be more immune to the corresponding attacks and its robustness to the adversarial perturbations will be enhanced. 

In the main text, Fig. 4(a-b) plots the results of the adversarial training to defense the FGSM and PGD attacks for the 3D CNN. In order to obtain this figure, we use $5001$ legitimate samples and their corresponding $5001$ adversarial samples as the training set for retraining the network. After every epoch we calculate the accuracy of legitimate samples and adversarial samples, respectively. The loss is calculated on both legitimate and adversarial samples. We mention that the adversarial samples used for training are different in every epoch. In each epoch, we use the current model and legitimate samples to generate adversarial samples, using both legitimate and adversarial samples to train the model, and then in the next epoch, we generate new adversarial samples by the model with updated parameters. From the figure, it is clear that the accuracy for both the legitimate samples and adversarial samples increase as the number of epochs increase, and saturate at notable values larger than $0.96$. This demonstrates that the retrained classifier is indeed much more robust to the adversarial perturbations generated by the corresponding attacking methods. 

\diff{In Fig.~2(d) of the main text we present that the activation map of the Ising model classifier becomes much flatter after the adversarial training, which accords with the symmetry of calculating $M$. This results in the effectiveness of the adversarial training on identifying the phase transition point: In Fig.~4(c-d) in the main text we demonstrate the Ising model classifier's prediction curves on different $M$ and different spin configurations before and after adversarial training. Since we only consider the $30\times30$ lattice, we divide $M$ from $0$ to $1$ into $450$ intervals and in each interval we randomly generate $100$ samples. We generate samples according to $M$ instead of $T$, which is in order to elaborate the classifier's behaviours around the small range of the transition point. From the Fig.~4(c-d) in the main text we can see that the classifier presents similar predictions on samples with the same $M$, and the confidence of being the ferromagnetic phase goes down as $M$ decreases. This indicates that the classifier has learnt the correct classification rule according to the order parameter $M$. In Fig.~\ref{fig: transition_averaging} we show the relation between the confidence of being the ferromagnetic phase and $M$ by averaging $100$ different spin configurations on each $M$ interval. We can clearly see that after adversarial training, the prediction curve becomes sharper and smoother, and the transition point identified at the confidence threshold $P_{\text{ferro}}=0.5$ moves closer to the theoretical critical temperature $T_c$. This implies that the classifier can identify the phase transition point more precisely after adversarial training.}

\diff{Besides the Ising model, We show that adversarial training can also help the topological phase classifier to make better inference. From table \ref{table: NCC}, we find that the activation map of the sixth kernel on topologically trivial phase ($I_{\chi=0}$) becomes less correlated with topologically nontrivial ones ($I_{\chi=1,-2}$). In addition, the activation maps between $I_{\chi=1}$ and $I_{\chi=-2}$ also become less correlated. This indicates that the kernel begins to learn how to distinguish two topologically nontrivial phases. In Fig.~\ref{fig:attmap_Chiral}(g-h), we plot the sixth kernel's activation map on the legitimate and adversarial examples after the adversarial training. We can find that $I_{\text{adv}}$ is much less correlated with $I_{\chi=1}$, which can also be verified by table \ref{table: NCC}. All these results imply that the classifier learned a more robust convolutional kernel, which can better distinguish three different topological phases, to do inferring.}

\diff{The effectiveness of the adversarial training can be also derived from another view: although one cannot expect a universal defense strategy that is able to make the phase classifiers robust to all types of adversarial perturbations, the defense against certain adversarial attack can help the model better learn the physical principle behind the problem. To support our claim, we try the discrete attack on the Ising model classifier after the adversarial training against continuous attack (FGSM in this example). The result is shown in Fig. \ref{fig: onePixel_against_FGSM}. We generate adversarial examples based on the test set with the Monte Carlo method on 41 different temperatures. Each temperature has 250 samples. Under each temperature, we enumerate one pixel and two pixels' flip on all 250 samples and find the ratio of
samples that can make the model do misclassification. We find that after adversarial training, the refined classifier is more robust to discrete spins flip, and the peaks move closer to the phase transition point.}

\end{document}